\documentclass[aps,pra,twocolumn,floatfix,superscriptaddress,
showkeys,amsmath,amssymb,longbibliography]{revtex4-2}		
 
\usepackage{braket}
\usepackage{graphicx}
\usepackage{dcolumn}
\usepackage{bm}
\usepackage[colorlinks,urlcolor=blue,citecolor=blue,linkcolor=blue]{hyperref}
\usepackage{subfig}
\usepackage[font=small,labelfont=bf,format=plain,justification=centerlast,labelsep=period]{caption}
\usepackage{tikz}
\usepackage{comment}
\usetikzlibrary{arrows,shapes,positioning,shadows,svg.path,backgrounds,fit}

\usepackage{verbatim}
\usepackage[normalem]{ulem}
\usepackage{orcidlink}

\begin{document}


\title{Quantum thermal machine as a rectifier}

\author{M. Santiago-Garc\'ia \orcidlink{0000-0002-6321-4853
}}
\affiliation{
Instituto Nacional de Astrof\'isica, \'Optica y Electr\'onica, Calle Luis Enrique Erro 1, Sta. Ma.  Tonantzintla, Puebla CP 72840, Mexico}
\author{O. Pusuluk\,\orcidlink{0000-0002-9167-7273
} }
\affiliation{Faculty of Engineering and Natural Sciences, Kadir Has University, 34083, Fatih, Istanbul, T\"{u}rkiye}
\author{\"O. E. M\"ustecapl{\i}o\u{g}lu\,\orcidlink{0000-0002-9134-3951
}}
\affiliation{Department of Physics, Ko\c{c} University, \.{I}stanbul, 
Sar{\i}yer, 34450, T\"urkiye}
\affiliation{TUBITAK Research Institute for Fundamental Sciences, 41470 Gebze, T\"urkiye}
\affiliation{Faculty of Engineering and Natural Sciences, 
Sabanci University, Tuzla, Istanbul, 34956, T\"urkiye}
\author{B. \c{C}akmak\,\orcidlink{0000-0002-6124-3925}} 
\affiliation{Department of Physics, Farmingdale State College—SUNY, Farmingdale, NY 11735, USA}
\author{R. Rom\'an-Ancheyta\,\orcidlink{0000-0001-6718-8587}}
\email{ancheyta@fata.unam.mx}
\affiliation{
Centro de F\'isica Aplicada y Tecnolog\'ia Avanzada, Universidad Nacional Aut\'onoma de M\'exico, Boulevard Juriquilla 3001, Querétaro 76230, Mexico
\\}

\date{\today}

\begin{abstract}

We study a chain of interacting individual quantum systems connected to heat baths at different temperatures on both ends. Starting with the two-system case, we thoroughly investigate the conditions for heat rectification (asymmetric heat conduction), compute thermal conductance, and generalize the results to longer chains. We find that heat rectification in the weak coupling regime can be independent of the chain length and that negative differential thermal conductance occurs. We also examine the relationship between heat rectification with entanglement and the entropy production. In the strong coupling regime, the system exhibits an asymmetric Rabi-type splitting in the thermal conductance, leading to enhanced heat transport and improved rectification inaccessible in the weak coupling. This setup represents the simplest quantum thermal machine that consumes incoherent resources and delivers entanglement while acting as a rectifier and heat valve.



\end{abstract}

\maketitle

\section{Introduction}

There has been increasing interest in investigating quantum heat transport at the nanoscale due to the development of quantum technologies and further efforts to miniaturize electronic, photonic, and thermal devices~\cite{Pekola_RMP_2021}. At this scale, one cannot ignore quantum effects, which, in turn, forced the development of a consistent framework to understand the thermodynamics of quantum systems~\cite{DeffnerCampbell}. Specifically, heat transport is usually modeled by coupled quantum systems in contact with different thermal baths at their edges, resulting in heat currents through the central system. Such settings are known as nonequilibrium boundary-driven quantum systems and are a topic of great interest in quantum thermodynamics and condensed matter physics~\cite{Landi_RMP_2022}.

One particularly compelling topic in quantum heat transport is thermal rectification. Like electronics seek to actively regulate electrical currents, the potential to fully control heat flow in nanodevices is a fascinating goal~\cite{Ben_2016,control_heatcurrent}, making thermal rectification a crucial area of research~\cite{Review_Malik_2022, Review_Liu_2019}. Devices that harness this effect are called thermal rectifiers or thermal diodes~\cite{Review_Wong_2021}. These devices allow heat to flow in one direction while decreasing (even blocking) it in the reverse direction when the temperatures of the baths are swapped, resulting in asymmetric heat conduction. Although Starr made the first observation of thermal rectification in 1935~\cite{Context_Rect}, significant progress in understanding and utilizing this phenomenon at the nanoscale has occurred in recent years, with studies conducted both theoretically~\cite{Teo_Rect1,Segal_PRL_2005,PRL_Segal_2009,Palafox_PRE_2022,PRR_Haack_2023} and experimentally~\cite{Exp_Rect5, Review_Zhao_2022, Exp_Rect1, Exp_Rect2, Exp_Rect3, Exp_Rect4}. From a practical point of view, these devices are intended to be compact, preferably with a specific thermal rectification behavior. 

In this paper, we study a chain of interacting qubits positioned between two thermal reservoirs at different temperatures, i.e., the chain serves as a bridge connecting hot and cold heat baths. This setup is one of the simplest quantum thermal machine (QTM)~\cite{Nicole_2024}, such that it consumes incoherent resources and delivers quantum coherence and entanglement~\cite{Bohr_Brask_2015, NJP_Geraldine_2020, Kenza}. We thoroughly analyze its thermal rectification behavior in the weak and strong coupling regimes and compare it to other setups with or without quantum entanglement to explore its advantages. Our research reveals that thermal rectification is the same in the weak coupling regime regardless of whether the bridge is a single-qubit (individual atom junction) or a multi-qubit chain, but not in strong coupling. Therefore, we analyze the thermal conductance in both regimes, enabling a more meaningful comparison between these setups. We show that achieving a negative differential thermal conductance (resistance) in the weak coupling regime is possible, a result not commonly observed, while an asymmetric Rabi-type splitting emerges in the strong coupling. 

We organize this article as follows. Sec.~\ref{sec:model} introduces the physical model we use to present the local master equation (ME), also known as phenomenological ME, based on the collision model approach of open quantum systems. In Sec.~\ref{sec:LME}, we conduct a comprehensive analysis and obtain a general analytical expression for the steady-state heat current of two weakly interacting qubits connecting two thermal baths at different temperatures. We analyze this setup's heat rectification behavior and thermal conductance. We examine the relation between heat rectification, entanglement, and the irreversible entropy production. In Sec.~\ref{sec:GME},  we extend our analysis to a setup in the strong coupling regime using the global master equation (GME), also known as microscopic ME. We conclude in Sec.~\ref{sec:conc}.

\section{The Model}\label{sec:model} 

In this and the following section, we use the highly versatile tool of quantum collision models~\cite{CM, Campbell2021}, also known as repeated interactions, to describe the behavior of our QTM. The basic idea in such models is to represent the environment as a set of \textit{ancillary} units that the target system of interest interacts with for a short time, $\tau$. The ancilla is discarded after interaction with the system, and a new one is brought into the picture. The dynamic of the target system is then described by the consecutive iterations of the steps summarized above. Taking the continuous time limit of this discrete-time (stroboscopic) evolution, one arrives at the well-known Lindblad master equation of open quantum systems~\cite{PRA_Ancheyta_2021}. 

\begin{figure}[t]
\centering
\includegraphics[width=1.0\columnwidth]{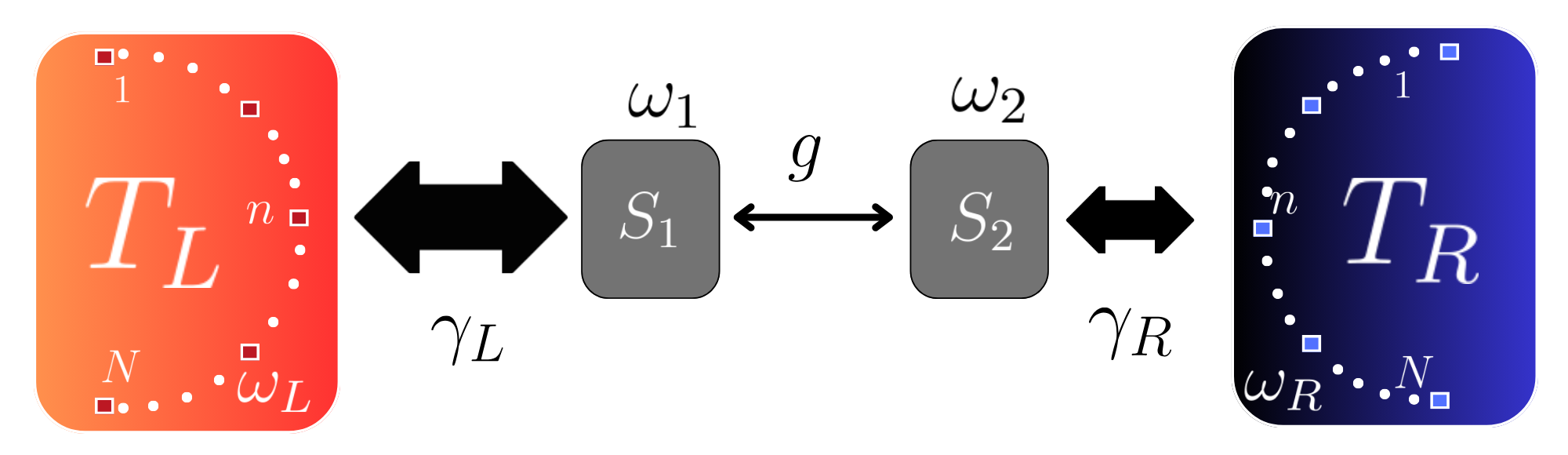}
\caption{Schematic representation of a quantum thermal machine (an entanglement engine) acting as a thermal rectifier described by the quantum collision model. Here, $\gamma_\lambda^{}$ is the coupling of each subsystem $S_{1,2}^{}$ with the heat bath $\lambda=L, R$ having the apparent temperature $T_\lambda$, and $g$ is the coupling strength between $S_1$ and $S_2$.}
\label{fig_setup_two_qubit}
\end{figure}

In particular, throughout this work, our primary target system consists of two interacting subsystems, each coupled with their heat reservoirs at different apparent temperatures~\cite{Latune_2019} as depicted in Fig.~\ref{fig_setup_two_qubit}. The left (hot) and right (cold) reservoirs are each represented by $N$ identical units, prepared in the Gibbs (thermal) state at the temperature of the corresponding bath. All interactions between the ancillas and subsystems have the same duration, and in between these collisions, an interaction between two subsystems takes place. This process is repeated consecutively. We can now explicitly present the mathematical details of our physical model outlined above. 

The quantum Hamiltonian describing the total system is given by
$ H=H_S^{}+H_{S_1S_2}^{}+H_B+H_{S_1{\rm L}}^{}+H_{S_2{\rm R}}^{}$, where $H_S^{}=\sum_{j=1}^2 H_{S_j}=\sum_{j=1}^2\hbar\omega_j^{}\, s_{j}^{\dagger} s_{j}^{}$ denote the bare Hamiltonians of the subsystems, ${H}_{S_1S_2}=g (s_{1}^{\dagger}s_{2}^{}+s_{1}^{}s_{2}^{\dagger})$ is the interaction between the them. $H_B=\sum_{\lambda}H_{\lambda}=\sum_{\lambda}\hbar\omega_\lambda^{}b_{\lambda}^{\dagger}b_{\lambda}^{}$ are the free Hamiltonians of the bath elements (ancilla), and ${H}_{S_j\lambda}^{}=\sqrt{\gamma_\lambda^{}}(s_{j}^{}b_{\lambda}^{\dagger}+s_{j}^{\dagger}b_{\lambda}^{})$ represents the system-bath interaction with $\lambda=L,R$ being the corresponding bath label (left and right) and should be scaled by $1/\sqrt{\tau}$ as detailed in \cite{PRA_Ancheyta_2021}, to avoid divergence in the continuous-time limit when $\tau \to 0$. Here, $\omega_j^{}$ and $\omega_\lambda^{}$ are the natural frequency of the subsystems and bath elements, respectively, $g$ is the coupling strength between subsystems, $\gamma_\lambda^{}$ denote the system-bath interaction strength, and $s_j^{\dagger}$ ($s_j^{}$) and $b_{\lambda}^{\dagger}$ ($b_{\lambda}^{}$) represent the creation (annihilation) quantum operators for the subsystems and bath elements, respectively. We assume that the frequency of each subsystem and the bath elements are the same (resonant conditions) $\omega_j^{}=\omega_\lambda^{}\equiv\omega$. This assumption, along with the energy-preserving form of the system-bath interaction, implies that any change in the energy of the central system is unambiguously due to the energy flow to or from the bath elements in the form of heat, not in the form of work~\cite{NJP_Chiara_2018}. Throughout this work, we set $\hbar=1$ and $k_{B}=1$ for simplicity.

At this point, we highlight an important advantage of employing the collision model approach in the current context. It allows us to examine scenarios where the central system and the bath elements may represent particles satisfying different quantum statistics in their average number of excitations by judiciously choosing the commutation or anti-commutation relations fulfilled by the elements forming the physical model. To this end, as shown below, we express the operators' commutation relations in the most general form~\cite{PRE_Sargsyan_2021}
\begin{equation}\label{b_commutation}
     s_j^{}s_{j}^{\dagger}-\epsilon_j^{}\,s_{j}^{\dagger}s_j^{}=1 \quad \textrm{and} \quad b_{\lambda}^{}b_{\lambda}^{\dagger}-\epsilon_{\lambda}^{}\,b_{\lambda}^{\dagger}b_{\lambda}^{}=1,
\end{equation}
where $\epsilon_{\lambda}^{} (\epsilon_{j}^{})=1$ for bosonic baths (subsystems), e.g., harmonic oscillators and $\epsilon_{\lambda}^{}(\epsilon_{j}^{})=-1$ for two-level systems (TLS), i.e., qubits, forming the reservoirs (subsystems).

Before presenting our results, we emphasize the existence of two distinct regimes available in the present setting, namely the weak and strong coupling limits. While the former assumes that the interaction $g$ is less than or similar to that of the interaction strength between a subsystem and its bath, i.e., $g\lesssim\gamma_{\lambda}$, the latter investigates the exact opposite limit, i.e., $g\gg\gamma_{\lambda}$. We will work within the continuous-time limit on both regimes and employ the corresponding master equation to describe the dynamics. However, the considerations leading to a thermodynamically consistent master equation in the weak and strong coupling regimes are different and must be handled with due care~\cite{Hofer, Global, Cattaneo}. 


\section{Weak Coupling: local approach}\label{sec:LME}

We start our discussion by presenting the corresponding master equation necessary to compute the heat current through the weakly interacting subsystems (our bridge). It is well-known that in the continuous-time limit, the dynamics of the target system of interest can be cast into the so-called local master equation (LME) in Gorini-Kossakowski-Sudarshan-Lindblad (GKLS) from~\cite{Global}. The Markovian LME for the density matrix $\rho$ of the central system is~\cite{NJP_Chiara_2018, AVS_Chiara_2022}
\begin{equation}\label{m_equation}
\begin{split}
&\frac{d \rho}{dt}\!=\!-i\big[H_{S_1}\!^{}+H_{S_2}\!^{}+H_{S_1S_2}^{}\,,\,\rho\big]\\ 
&\hspace{0.28cm}+\gamma_L^{-}\mathcal{D}[s_1^{}]\rho+\gamma_L^{+}\mathcal{D}[s_1^{\dagger}]\rho+\gamma_R^{-}\mathcal{D}[s_2^{}]\rho + \gamma_R^{+}\mathcal{D}[s_2^{\dagger}]\rho,
\end{split}
\end{equation}
where $\mathcal{D}[x]\rho\equiv x\rho x^{\dagger}-\frac{1}{2}(x^{\dagger}x\rho+\rho x^{\dagger}x)$ are the Lindblad super-operators (dissipators) acting {locally} on the subsystems. The dissipation rates $\gamma_{\lambda}^{+}\equiv\gamma_{\lambda}^{}\langle b_{\lambda}^{\dagger} b_{\lambda}^{}\rangle=\gamma_{\lambda}^{}n_{\lambda}^{}$ and $\gamma_{\lambda}^{-}\equiv\gamma_{\lambda}^{}\langle b_{\lambda}^{} b_{\lambda}^{\dagger}\rangle=\gamma_{\lambda}^{}(1+\epsilon_\lambda^{} n_{\lambda}^{})$ satisfy the detailed balance condition. Since each bath element is in the thermal state $\rho_\lambda^{\rm th}=\exp(-H_\lambda^{}/k_B^{}T_\lambda)Z_\lambda^{-1}$, with $Z_\lambda^{}$ the partition function, we can express the average number of excitations as a function of the temperature $T_\lambda^{}$ in the general form~\cite{PRE_Sargsyan_2021} 
\begin{equation}\label{n_distribution}
    n_{\lambda}^{}=\big[{\exp(\hbar\omega_\lambda^{}/k_B^{}T_{\lambda}^{})-\epsilon_{\lambda}^{}}\big]^{-1}.
\end{equation}
For $\epsilon_{\lambda}^{}=1$, $n_\lambda^{}$ describes the Bose-Einstein distribution, while for $\epsilon_{\lambda}^{}=-1$, Eq.~(\ref{n_distribution}) is the Fermi-Dirac distribution.

\subsection{Heat current}\label{heat_subsection}

As we assumed resonant conditions (where the work cost in each collision is strictly zero), the whole energy exchange between the central system and the corresponding thermal baths is in the form of heat. Then, following the standard definition~\cite{NJP_Geraldine_2020, Alicki2018}, we can calculate the heat flow $Q_\lambda^{}$ from the bath $\lambda$ to the corresponding subsystem $S_j^{}$ as $Q_{\lambda}^{}={\rm tr}\{H_{S}^{}(\gamma_{\lambda}^{-}\mathcal{D}[s_{j}^{}]\rho + \gamma_{\lambda}^{+}\mathcal{D}[s_{j}^{\dagger}]\rho)\}$.
Since ${\rm tr}\{s_{j}^{\dagger}s_{j}^{}\, \rho\}=\langle s_{j}^{\dagger}s_{j}^{}\rangle$, we specifically have
\begin{equation}
    Q_{\lambda}^{}=\omega \gamma_{\lambda}^{}\big(n_{\lambda}^{}-[1+(\epsilon_{\lambda}^{}-\epsilon_{j})n_{\lambda}^{}]\langle s_{j}^{\dagger}s_{j}^{}\rangle\big).
\end{equation}

Focusing on the steady state (limit cycle) of the QTM, which we denote by the label ``SS" in all physical quantities, the heat current from the hot bath to the cold bath is defined by $J_{\rm SS}^{(2)}\equiv Q_{L}^{\rm SS}-Q_{R}^{\rm SS}$~\cite{NJP_Geraldine_2020}, where the superscript $(2)$ denotes the number of subsystems making up the central system. In the steady state, the heat in and out of the central system must be the same, i.e., $Q_{L}^{\rm SS}=-Q_{R}^{\rm SS}$ and thus the heat current can be written in terms of a single expectation value, for instance $\langle s_{1}^{\dagger}s_{1}^{}\rangle_{\rm SS}^{}$. Without loss of generality, in the remainder of this work, we consider the interacting subsystems making up the central system to be of the same nature in terms of particle exchange statistics, i.e., satisfy the same commutation/anti-commutation relations $\epsilon_1^{}=\epsilon_2^{}\equiv\epsilon_a^{}$. Following the detailed calculations we present in Appendix~\ref{appendix:a}, it is possible to determine the aforementioned expectation values in an explicit closed form and write the heat current as
\begin{equation}\label{general_heat_2}
    J_{\rm SS}^{(2)}=\alpha\, J_{\rm SS}^{(1)},
\end{equation}
where the scaling factor
\begin{equation}\label{alpha_factor}
    \alpha=\Big(1+\frac{\gamma_L^{}\gamma_R^{}}{4g^2}\big[1+(\epsilon_{L}^{}-\epsilon_{a}^{})n_{L}\big]\big[1+(\epsilon_{R}^{}-\epsilon_{a}^{})n_{R}^{}\big]\Big)^{-1}
\end{equation}
and the heat current through a single central system is
\begin{equation}
    J_{\rm SS}^{(1)}=\frac{2\omega\gamma_{L}^{}\gamma_{R}^{}\left[(n_{L}^{}-n_{R}^{})+(\epsilon_{R}^{}-\epsilon_{L}^{})n_{L}^{}n_{R}^{}\right]}{\gamma_{L}^{}+\gamma_{R}^{}+\gamma_{L}^{}(\epsilon_{L}^{}-\epsilon_{a}^{})n_{L}^{}+\gamma_{R}^{}(\epsilon_{R}^{}-\epsilon_{a}^{})n_{R}^{}}.
\end{equation}

Equation~(\ref{general_heat_2}), our first main result, is an unreported analytical expression and clearly shows that $J_{\rm SS}^{(2)}$ is proportional to $J_{\rm SS}^{(1)}$, hence Eq.~(\ref{general_heat_2}) generalizes all the results on heat transport and rectification of the pivotal single-atom junction from our recent work~\cite{Palafox_PRE_2022}. We also recover the case of a QTM made of bosonic reservoirs ($\epsilon_\lambda^{}=1$) and oscillators (qubits) subsystems $\epsilon_a^{}=\pm 1$~\cite{AVS_Chiara_2022, PRR_Haack_2023}. The interaction strength between the subsystems, $g$, appears only in the $\alpha$ factor, and when $g\rightarrow 0$, $J_{\rm SS}^{(2)}\rightarrow 0$. It is instructive to look at some other limiting cases of the $\alpha$ factor to understand the behavior of $J_{\rm SS}^{(2)}$. In general, $\alpha$ is bounded between $0$ and $1$, which implies that $J_{\rm SS}^{(2)}\leq J_{\rm SS}^{(1)}$ for any choice of parameters. In the case of $\gamma_L^{}\gamma_R^{}/4g^{2}\ll 1$, $\alpha\rightarrow 1$ and $J_{\rm SS}^{(2)}\approx J_{\rm SS}^{(1)}$, which means that the two subsystems are strongly coupled and the ``center of mass'' only interacts with the heat baths~\cite{AVS_Chiara_2022}; however, this interpretation may fall outside the validity range of the LME. In the low-temperature limit $k_B^{}T_\lambda^{}/\hbar\omega_\lambda\ll 1$, $n_\lambda^{}$ vanishes; therefore, $\alpha$ becomes independent of the $\epsilon_{\lambda}^{}$ value, i.e., the exchange statistics of the elements making up the $\lambda$ bath. We complete the analysis of how $\alpha$ depends on different model parameters in Appendix~\ref{appendix:b}.

We can compute the generated coherence of the two interacting subsystems forming the bridge as follows: $\mathcal{C} {\rm}=\langle s_{1}^{}s_{2}^{\dagger}-s_{1}^{\dagger}s_{2}^{}\rangle/2$. With the help of (\ref{S1S1EqMot}), we verify that $\mathcal{C}_{\rm SS}^{}$ is a purely imaginary complex number given by $ \mathcal{C}_{\rm SS}^{}=\frac{i}{4g\omega}J_{\rm SS}^{(2)}$, providing a clear connection between heat current and system coherence. Such a relationship implies that a non-zero heat current generates non-zero coherence. Although this observation was reported in~\cite{Manzano, NJP_Geraldine_2020} for the particular case of coupled central qubits and bosonic baths, here we demonstrate its general validity for any combinations of $\epsilon_\lambda^{}$ and $\epsilon_a^{}$, including hybrid quantum structures, i.e., $\epsilon_L^{}\neq\epsilon_R^{}$~\cite{PRL_Segal_2009}. Since $\langle s_{1}^{}s_{2}^{\dagger}\rangle_{\rm SS}^{}$ is the complex conjugate of $\langle s_{1}^{\dagger}s_{2}^{}\rangle_{\rm SS}^{}$, one can also get the heat current from the alternative expression $J_{\rm SS}^{(2)}=-4\omega g{\rm Im}\{\langle s_{1}^{\dagger}s_{2}\rangle_{\rm SS}^{}\}$~\cite{Manzano}.

\subsection{Thermal Rectification}

To assess the thermal rectification performance of our setup, we use the figure of merit known as {\it contrast}, defined as $C_{\rm SS}^{(2)}\!=\!\big|({J_{\rm SS}^{(2)\rightarrow}\!+\!J_{\rm SS}^{(2)\leftarrow}})/({J_{\rm SS}^{(2)\rightarrow}\!-\!J_{\rm SS}^{(2)\leftarrow}})\big|$~\cite{Landi_RMP_2022}.
Here, $J_{\rm SS}^{(2)\rightarrow}$ denotes the heat current of Eq.~(\ref{general_heat_2}), and $J_{\rm SS}^{(2)\leftarrow}$ is the heat current of Eq.~(\ref{general_heat_2}) when the temperatures of the baths are interchanged, i.e., $T_L \leftrightharpoons T_R$, while keeping all other parameters the same. When $C_{\rm SS}^{(2)}=0$, there is symmetric heat conduction through the system, indicating the absence of thermal rectification. Conversely, if $C_{\rm SS}^{(2)}=1$, heat rectification is at its maximum, and the system is considered a perfect thermal rectifier or perfect thermal diode, where heat can flow in one direction but is completely blocked in the other. Exploiting the relationship $J_{\rm SS}^{(2)}=\alpha\, J_{\rm SS}^{(1)}$, we can write $C_{\rm SS}^{(2)}$ in terms of the heat current through a single central system:
\begin{equation}\label{contrast_general}
     C_{\rm SS}^{(2)}=\left|\frac{A J_{\rm SS}^{(1)\rightarrow}+B J_{\rm SS}^{(1)\leftarrow}}{A J_{\rm SS}^{(1)\rightarrow}-B J_{\rm SS}^{(1)\leftarrow}}\right|,
\end{equation}
where A and B are given by
\begin{align}\label{AandB}
    A&=1\!+\!\frac{\gamma_{L}^{}\gamma_{R}^{}}{4g^2}\big[1\!+\!(\epsilon_{L}^{}\!-\!\epsilon_{a}^{})n_L^{}(T_R)\big]\big[1\!+\!(\epsilon_{R}^{}\!-\!\epsilon_{a}^{})n_R^{}(T_L)\big],\nonumber\\
    B&=1\!+\!\frac{\gamma_{L}^{}\gamma_{R}^{}}{4g^2}\big[1\!+\!(\epsilon_{L}^{}\!-\!\epsilon_{a}^{})n_L^{}(T_L)\big]\big[1\!+\!(\epsilon_{R}^{}\!-\!\epsilon_{a}^{})n_R^{}(T_R)\big].
\end{align}
Due to the exchange operation of the bath temperatures, $n_\lambda^{}(T_x)$ in Eq.~(\ref{AandB}) represents the distribution described by Eq.~(\ref{n_distribution}), with the temperature $T_\lambda$ replaced by $T_x$. We note that although $J_{\rm SS}^{(2)}$ is proportional to $J_{\rm SS}^{(1)}$, this does not imply, in general, that $C_{\rm SS}^{(2)}$ should be equal to $C_{\rm SS}^{(1)}$. Indeed, this equality occurs only when the baths are of the same nature ($\epsilon_L^{}=\epsilon_R^{}$), such that $n_L^{}(T_R)=n_R^{}(T_R)$, $n_R^{}(T_L)=n_L^{}(T_L)$, and $A=B$. In this scenario, $C_{\rm SS}^{(2)}=C_{\rm SS}^{(1)}$ regardless of the value of $\epsilon_a^{}$. Consequently, the setup depicted in Fig.~\ref{fig_setup_two_qubit} behaves as if it were a single central system in terms of rectification, and its performance is independent of the coupling strength $g$ between the interacting subsystems forming the bridge. 

When the left bath shares the same statistic as the central system, but differs from the right bath, we have $\epsilon_{L}^{}=\epsilon_{a}^{}=-\epsilon_{R}^{}$. In this case, $A=1+\gamma_L^{}\gamma_R^{}[1+2\epsilon_R^{}n_R^{}(T_L)]/4g^2$ and $B=1+\gamma_L^{}\gamma_R^{}[1+2\epsilon_R^{}n_R^{}(T_R)]/4g^2$. If $\epsilon_R^{}=\pm 1$, then $1\pm 2\,n_R^{}(T_x)=[\coth(\hbar\omega/2k_B^{}T_x)]^{\pm1}$. Similar expressions hold for the case where $-\epsilon_L^{}=\epsilon_a^{}=\epsilon_R^{}$. Thus, across the four possible hybrid quantum structures, where the baths exhibit different statistical properties ($\epsilon_L^{}\neq\epsilon_R^{}$), we find that $C_{\rm SS}^{(2)}\neq C_{\rm SS}^{(1)}$.

In what follows, we will discuss our results for the particular case where the central system is made of two coupled qubits, and both thermal baths have a bosonic nature, i.e., harmonic oscillators. The reason why we choose this particular setting is due to the facts that entanglement is easier to quantify for a pair of qubits and in the strong-coupling regime that we will analyze in Sec~\ref{sec:GME}, the global master equation describing the dynamics is derived only for the aforementioned setting. For, $\epsilon_{a}=-1$ and $\epsilon_{\lambda}=1$, difference is $\epsilon_\lambda^{}-\epsilon_a^{}=2$ and the steady-state heat current (\ref{general_heat_2}) reduces to
\begin{equation}\label{heat_current_2}
J_{\rm SS}^{(2)}\!=\!\Big[1\!+\!\frac{\gamma_{L}^{}\gamma_{R}^{}}{4g^2}\coth\!{\Big(\frac{\hbar\omega}{2k_B^{}T_{L}}\Big)}\coth\!{\Big(\frac{\hbar\omega}{2k_B^{}T_{R}}\Big)}\Big]^{-1}\times J_{\rm SS}^{(1)},
\end{equation}
where 
\begin{equation}\label{Conductance_1_qubit}
    J_{\rm SS}^{(1)}=\frac{2\omega\gamma_{L}^{}\gamma_{R}^{}(n_L^{}-n_R^{})}{\gamma_{L}^{}(1+2n_L^{})+\gamma_{R}^{}(1+2n_R^{})},
\end{equation}
$n_\lambda^{}$ is obtained from Eq.~(\ref{n_distribution}) with $\epsilon_\lambda^{}=1$. In this case $A=B$, $C_{\rm SS}^{(2)}=C_{\rm SS}^{(1)}$, and
\begin{equation}\label{CBQB}
    C_{\rm SS}^{(1)}=\left|\frac{\gamma_L^{}-\gamma_R^{}}{\gamma_L^{}+\gamma_R^{}}\right|\left|\frac{\coth{(\omega/2T_R)}-\coth{(\omega/2T_L)}}{\coth{(\omega/2T_R)}+\coth{(\omega/2T_L)}}\right|. 
\end{equation}
Note that $C_{\rm SS}^{(1)}=0$ if the coupling to the baths is symmetric, i.e., $\gamma_{L}^{}=\gamma_R^{}$, and hence it is a necessity to have parametric asymmetry in the bath coupling parameters ($\gamma_L^{}\neq\gamma_R^{}$) for thermal rectification. 

Recall that the current setup is a QTM capable of generating entanglement between the system qubits, as shown in several works~\cite{Bohr_Brask_2015,NJP_Geraldine_2020}. It is also possible to prove that a minimum current, $J_{c}$, is required to generate entanglement in the steady state,  irrespective of the nature of the baths. This critical heat current, which acts as an entanglement witness, is given by $J_{c} =4g\omega \sqrt{p_{1}p_{4}}$, where $p_{1}$ and $p_{4}$ are the populations of the doubly excited and ground state of the two-qubit density matrix, respectively. The details and the definitions of $p_{1}$ and $p_{4}$ in terms of other model parameters can be found in Appendix \ref{appendix:c}. Remarkably, we show that even when $\gamma_{L}^{}=\gamma_{R}^{}$, a minimum heat current can still generate entanglement, indicating that a QTM can produce entanglement without displaying heat rectification. This result suggests no fundamental relation between heat rectification and entanglement generation but a practical one. For instance, we may operate the QTM at the critical point of $J_c$ with $\gamma_L^{}\neq\gamma_R^{}$; once we swap the temperatures, the resulting heat current can drop below $J_c$, extinguishing the entanglement. On the other hand, in Appendix~\ref{appendix:d}, we show that thermal rectification is always tied to a change in the entropy production rate~\cite{Landi_RMP_2021} when the temperatures are swapped.

\subsection{Thermal conductance}

Since heat rectification is identical for the two-qubit and single-qubit configurations as the central system, we focus on analyzing and comparing their thermal conductance. Specifically, we examine the thermal conductance in the linear regime, defined as $\kappa_{\rm SS}^{(2)}\equiv\lim_{\Delta T \to 0}^{}{J_{\rm SS}^{(2)}}/\Delta T$~\cite{Conductance,Kondo,Yamamoto_2021}. By substituting the heat current expression from Eq.~(\ref{general_heat_2}), we find $\kappa_{\rm SS}^{(2)}=\alpha(T)^{}\kappa_{\rm SS}^{(1)}$, where $\alpha(T)$ denotes the scaling factor of Eq.~(\ref{alpha_factor}) evaluated at $T_L=T_R=T$, and $\kappa_{\rm SS}^{(1)}$ is the general linear thermal conductance for a single central system acting as the bridge given by
\begin{equation}
\kappa_{\rm SS}^{(1)}=\frac{\omega^{2}\gamma_{L}^{}\gamma_{R}^{}\, n_L^{}n_a^{}n_R^{}\exp(\hbar\omega/k_B^{}T)}{\left(\gamma_L^{}n_L^{}+\gamma_R^{}n_R^{}\right)k_B^{}T^{2}}.
\end{equation}
For the specific case of interest involving two qubits coupled to bosonic baths with $\epsilon_{a}^{}=-1$ and $\epsilon_{\lambda}^{}=1$, we get
\begin{equation}\label{kappa_ss_2}
    \kappa_{\rm SS}^{(2)}=\Big[1+\frac{\gamma_{L}^{}\gamma_{R}^{}}{4g^2}\coth{^{2}\Big(\frac{\hbar\omega}{2k_B^{}T}\Big)}\Big]^{-1} \kappa_{\rm SS}^{(1)},
\end{equation}
and $\kappa_{\rm SS}^{(1)}=\omega^{2}{\gamma_{L}^{}\gamma_{R}^{}}{\rm csch}\left(\hbar\omega/k_B^{}T\right)/[(\gamma_{L}^{}+\gamma_{R}^{}){k_BT^{2}}]$~\cite{Palafox_PRE_2022}. As expected, if $\gamma_{L}^{}\gamma_{R}^{}/4g^2\ll 1$, then $\kappa_{\rm SS}^{(2)} \to \kappa_{\rm SS}^{(1)}$ and $\kappa_{\rm SS}^{(1)}$ represents an upper bound for $\kappa_{\rm SS}^{(2)}$ in the local approach. 

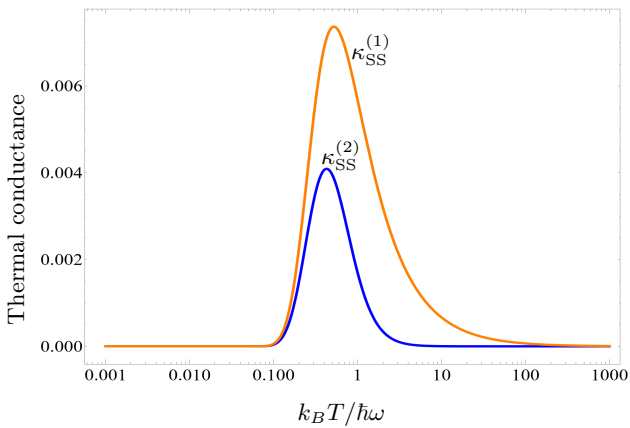
\begin{figure}[b]
\centering
\begin{tikzpicture} 
  \node (img1) {\includegraphics[height=5cm, width=7.8cm]{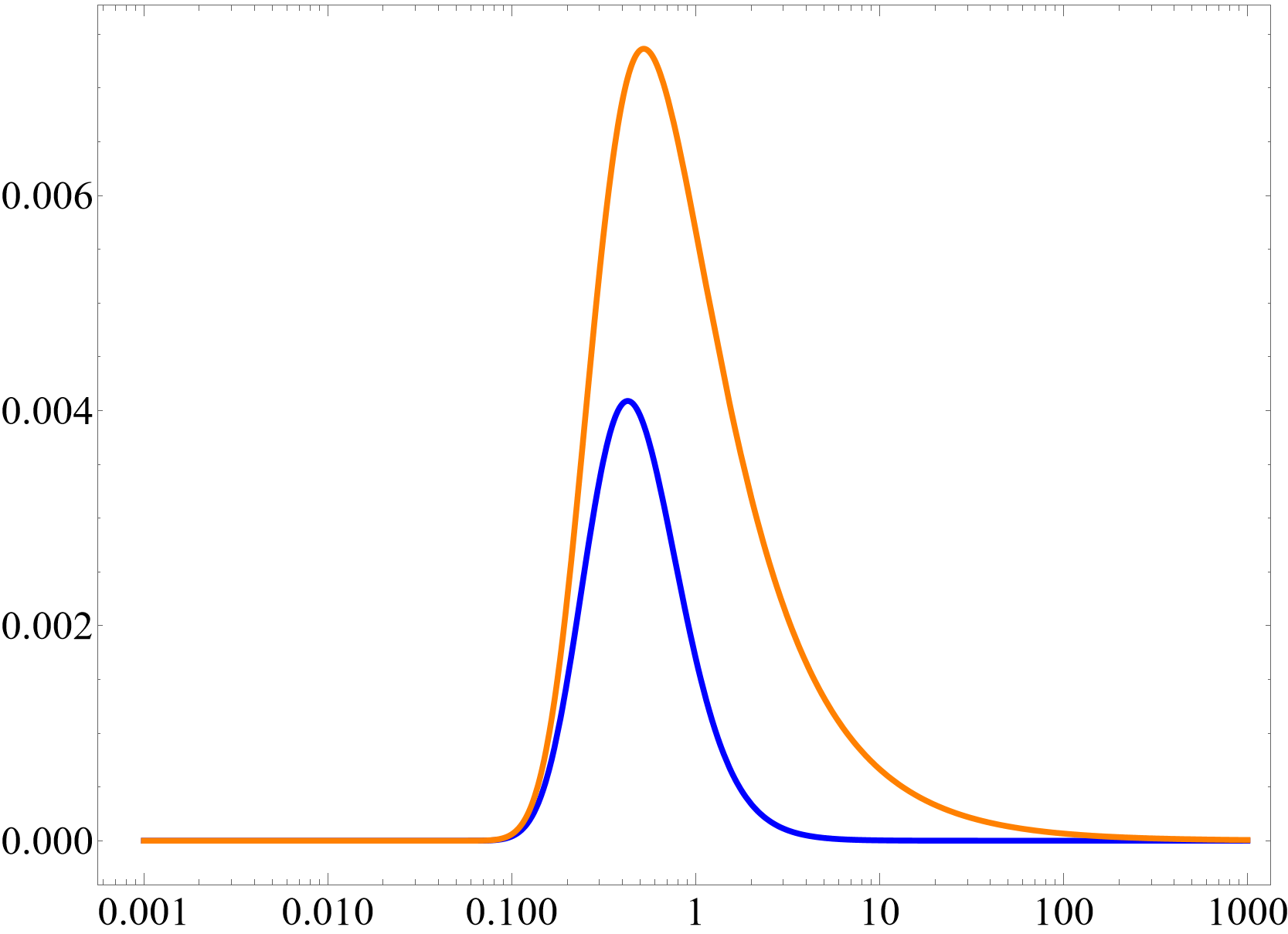}};
  \node[above=of img1, node distance=0cm, yshift=-6.8cm,xshift=0.1cm] {$k_{B}T/\hbar\omega$};
  \node[above=of img1, node distance=0cm, yshift=-3.4cm,xshift=0.1cm] {$\kappa_{\rm SS}^{(2)}$};
  \node[above=of img1, node distance=0cm, yshift=-2.0cm,xshift=0.5cm] {$\kappa_{\rm SS}^{(1)}$};
  \node[above=of img1, node distance=0cm,rotate=90, yshift=4cm,xshift=-3.8cm] {{\color{black} Thermal conductance}};
\end{tikzpicture}
\caption{Thermal conductance. The blue line represents $\kappa_{\rm SS}^{(2)}$ [see Eq.~(\ref{kappa_ss_2})], while the orange curve corresponds to $\kappa_{\rm SS}^{(1)}$. We set the parameters $\gamma_{L}^{}=0.01\omega$, $\gamma_{R}^{}=0.02\omega$, and $g=0.01\omega$.}\label{weak_conductance}
\end{figure}

Figure~\ref{weak_conductance} illustrates the temperature dependence of the linear thermal conductance, which exhibits Schottky-type behavior~\cite{Yamamoto_2021}. This behavior is characterized by the suppression of thermal conductance at both low and high temperatures, with a single peak around $T\approx \hbar\omega/2k_B^{}$. A similar but experimental Schottky-type behavior is observed in the specific heat of a two-level system as a function of temperature~\cite{Specific_heat, Schot_Exp}. The maximum value of Eq.~(\ref{kappa_ss_2}) can be well approximated as 
\begin{equation}\label{kappa_ss_2_max}
\kappa_{\rm SS_{Max}}^{(2)}\approx \Big[1+\frac{\gamma_{L}\gamma_{R}}{4g^2}\coth{^{2}(1)}\Big]^{-1}\kappa_{\rm SS_{Max}}^{(1)},
\end{equation}
where $    \kappa_{\rm SS_{Max}}^{(1)}\approx 4 {\rm csch}(2) {\gamma_{L}^{}\gamma_{R}^{}}/(\gamma_{L}^{}+\gamma_{R}^{})$. It is noteworthy that $\kappa_{\rm SS_{Max}}^{(2)}$ and $\kappa_{\rm SS_{Max}}^{(1)}$ are independent of $\omega$. While changing 
$\omega$ alters the location of the maximum value, it does not affect its magnitude. Hence, if a specific temperature regime is targeted and maximum conductance is desired within that regime, we will only need to modify the qubit frequency. Consequently,  $\kappa_{\rm SS_{Max}}^{(1)}$ ($\kappa_{\rm SS_{Max}}^{(2)}$) only depend on the coupling constants $\gamma_{L}^{}$, $\gamma_{R}^{}$ (and $g$).

Interestingly, this quantum thermal machine, composed of two coupled qubits, can exhibit negative differential thermal conductance (NDTC) in the weak coupling regime. NDTC arises when the heat current decreases as the temperature bias increases, i.e., when $\partial J_{\rm SS}^{(2)}/\partial\Delta T<0$, indicating  that
\begin{equation}
    \frac{\partial\alpha}{\partial\Delta T}J_{\rm SS}^{(1)}+\alpha\frac{\partial J_{\rm SS}^{(1)}}{\partial\Delta T}<0.
\end{equation}
However, $\partial J_{\rm SS}^{(1)}/\partial\Delta T$ and $\alpha$ remain positive for any combination of $\epsilon_\lambda^{}$ and $\epsilon_a^{}$, Therefore, for NDTC to occur, it is necessary that $\partial \alpha/\partial\Delta T<0$, which, in the limit $T_R^{}\rightarrow 0$, leads to the condition
\begin{equation}
T_{L}^{}> \ln\left[1+\frac{2\gamma_{L}^{}\sqrt{\gamma_{R}^{}/4g^2}}{\sqrt{(\gamma_{R}^{}+\gamma_{L}^{})(1+\gamma_{L}^{}\gamma_{R}^{}/4g^2)}}\right]^{-1}.
\end{equation}
This implies a suppression of the heat current as the temperature of the hot (left) bath increases. This result is significant because NDTC is an essential requirement for the operation of a thermal transistor~\cite{Li_NDTC, Miranda_Review_2017, PRR_Ozgur_202}, and is in contrast to previous studies in which NDTC achievement typically required strong coupling with thermal baths and nonlinear interactions~\cite{PRE_NDTC_Ren_2019,Segal_NDTC,Segal_NDTC1}.

\begin{figure}[t]
\centering
\includegraphics[width=.48\textwidth,keepaspectratio=true]{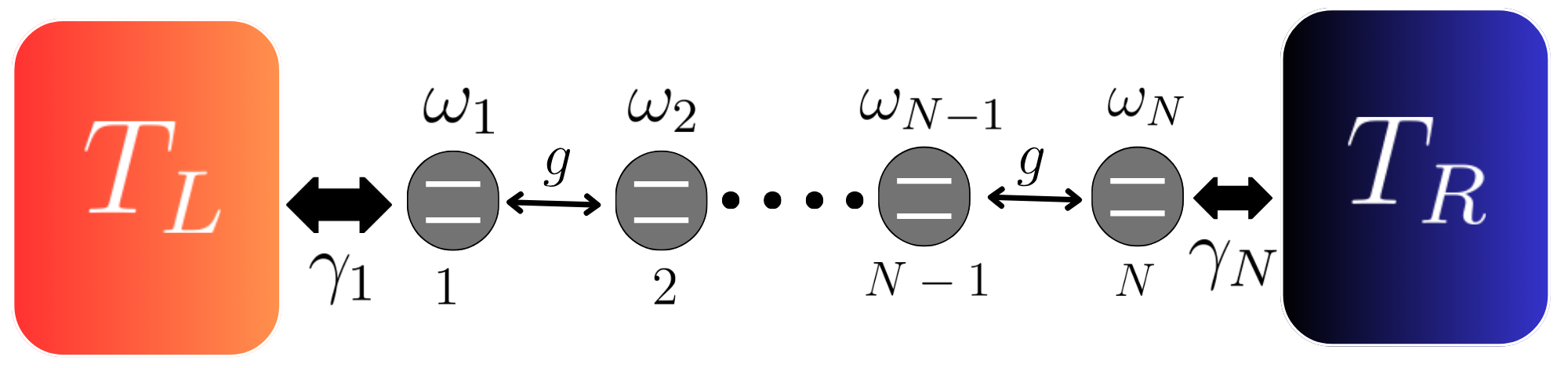}
\caption{Schematic representation of an N-qubit chain in contact with hot (left) and cold (right) thermal baths.}
\label{setupchain}
\end{figure}

\subsection{Generalization to N-qubit chain}

In Ref.~\cite{Manzano}, the authors found that for a chain of N-qubits with the same frequency, as illustrated in Fig.~\ref{setupchain}, the local master equation associated with the density matrix of the chain is $\dot{\rho}=-{i}[H_{\rm chain},\rho] + \mathcal{L}_1^{}\rho + \mathcal{L}_N^{}\rho$, where
\begin{equation}
H_{\rm chain}={\sum}_{k=1}^N\frac{\hbar\omega}{2}\sigma_k^z+{\sum}_{k=1}^N \hbar\, g (\sigma_k^+\sigma_{k+1}^-+\sigma_k^-\sigma_{k+1}^+).
\end{equation}
$\mathcal{L}_i^{}\rho=\gamma_{i}^{-}D[s_{i}^{}]\rho + \gamma_{i}^{\dagger}D[s_{i}^{\dagger}]\rho$ are the local Lindblad superoperators that describe the dissipation processes (absorption or emission) due to the coupling of the first ($i=1$) and last ($i=N$) qubits to thermal baths.

The steady-state heat current, $J_{\rm SS}^{(N)}$, as described in Eq.~(12) of~\cite{Manzano}, can be rewritten using our notation as 
\begin{equation}
\begin{split}
     J_{\rm SS}^{{(N)}}=&\frac{4g^{2} }{4g^{2}+\gamma_{1}^{}\gamma_{N}^{}\coth\big(\beta_L^{}\omega/2\big)\coth{ \big({\beta_{R}^{}\omega}/{2}\big)}}\\
     &\qquad\times \frac{2\omega \gamma_{1}^{}\gamma_{N}^{}(n_{L}^{}-n_{R}^{})}{\gamma_{1}^{}(1+2n_{L}^{})+\gamma_{N}^{}(1+2n_{R}^{})},
 \end{split}
\end{equation} 
where $\beta_\lambda^{}=(k_BT_\lambda)^{-1}$ is the inverse temperature. We observe that the above expression is equivalent to our Eq.~(\ref{heat_current_2}), i.e., we find that $J_{\rm SS}^{{(N)}}=J_{\rm SS}^{{(2)}}$. The heat current is independent of the size of the chain as long as $N\geq 2$. For $N=1$, we refer to Eq.~(\ref{Conductance_1_qubit}). This observation allows us to identify easily that the corresponding contrast will be $C_{\rm SS}^{(N)}=C_{\rm SS}^{(2)}=C_{\rm SS}^{(1)}$.
Therefore, a chain of N-qubits rectifies in the same manner as a setup with just one qubit acting as the central system~\cite{PRR_Haack_2023}. The first case discussed in the preceding section is just a particular case of a chain made of $N=2$ qubits. This result is significant from a practical point of view because it tells us there is no difference between using a long qubit chain or just one qubit as thermal rectifiers in the weak coupling case. Heat is difficult to focus or confine in small dimensions; in practice, it is challenging to separate a hot bath and a cold bath with only one or two qubits; therefore, using a chain is more attractive and feasible for practical implementations. Note that the heat current $J_{\rm SS}^{(N)}$ is independent of the chain's size, making a clear violation of Fourier's law of heat conduction~\cite{Manzano}.

By taking $T_{R} \to 0$, in Eq.~(\ref{CBQB}) and considering the high-temperature limit for the hot bath, i.e., $T_{L}\gg 1$, the contrast can be well approximated by $C_{\rm SS}^{(N)} \approx|(\gamma_1^{}-\gamma_N^{})/(\gamma_1^{}+\gamma_N^{})|$, which is independent of $\omega$ and $g$. Therefore, $C_{\rm SS}^{(N)}$ converges to a value that depends only on the ratio $\gamma_1^{}/\gamma_N^{}$. For the thermal conductance, it is easy to show that $\kappa_{\rm SS}^{(2)}=\kappa_{\rm SS}^{(N)}$, where $\kappa_{\rm SS}^{(N)}$ is the linear thermal conductance of the N-qubits chain. Remarkably, we find that this relation holds for a chain of N harmonic oscillators~\cite{Manzano1}. Thus, we can relate the thermal conductance of the three setups presented: $\kappa_{\rm SS}^{(N)}=\kappa_{\rm SS}^{(2)} \leq \kappa_{\rm SS}^{(1)}$.

All three setups studied above exhibit the same thermal rectification properties, meaning the setup choice cannot be based solely on rectification. Regarding thermal conductance, $\kappa_{\rm SS}^{(2)}$ and $\kappa_{\rm SS}^{(N)}$ are identical and are bounded by $\kappa_{\rm SS}^{(1)}$.  The single-qubit setup as the central system is the preferred choice for maximizing conductance. However, there is no distinction in conductance between the two-qubit chain and the N-qubit chain. From a practical perspective, the two-qubit setup emerges as the next best option. Moreover, if generating entanglement is a priority—something impossible with a single-qubit system—the two-qubit setup should be selected with appropriately tuned parameters.

It is important to emphasize that the above analysis assumes a local approach for both the two-qubit and N-qubit chains, which provides an accurate description of the system's dynamics under weak coupling between subsystems~\cite{NJP_Geraldine_2020,Global,Hofer,Cattaneo}. It is also worth noting that $J_{\rm SS}^{}$, $C_{\rm SS}^{}$, and $\kappa_{\rm SS}^{}$  could change if the coupling strength constant $g$ between the subsystems increases, as explored in the next section. This should not be confused with the coupling to the baths, $\gamma_{\lambda}^{}$, which will be small in all the cases discussed in this work.

\section{Strong coupling: global approach}\label{sec:GME}

When the coupling strength between the subsystems, $g$, is comparable to or exceeds the dissipation rates, $\gamma_\lambda^{}$, the validity of writing the dissipative part of the master equation as a sum of individual Lindblad superoperators acting locally on the subsystems becomes questionable~\cite{Hofer, Global}. Consequently, the master equation we used in previous sections may fail to describe the dynamics of our QTM in a physically consistent manner~\cite{Kosloff}. While doable in the strong coupling regime, it is too cumbersome to derive a global master equation when aiming to retain arbitrary particle statistics for the baths and subsystem elements. Therefore, to maintain clarity and conciseness in the presentation of results, we limit our use of the global master equation to the specific case of bosonic thermal baths and resonant coupled central qubits.

In the strong-coupling regime, the reservoirs couple to the global degrees of freedom of the composite system connecting them, and the jump operators drive transitions between eigenstates of the total Hamiltonian of both subsystems, i.e., $H_{S}+H_{S_1S_2}^{}$. For the particular case of two coupled qubits, the four eigenstates written in the single-qubit energy eigenbasis \{$ |00\rangle,  |01\rangle, |10\rangle, |11\rangle$\}, are  $|0\rangle\equiv|00\rangle$, $|\omega_{\pm}\rangle\equiv(|01\rangle\pm|10\rangle)/\sqrt{2}$, and $|2\rangle\equiv|11\rangle$. Following the treatment in~\cite{Global,Hofer, Cattaneo}, the GME is~\cite{NJP_Geraldine_2020}
\begin{equation}\label{Lindblad_GME}
\begin{split}
    &\dot\rho=-i[H_{S_1}+H_{S_2}+H_{S_1S_2}^{}\,,\,\rho]\\  
    &+{\sum}_{\varphi}\,\,\gamma_{L}^{-}(\omega_{\varphi})\mathcal{D}[\mathbb{L}_{1}^{}(\omega_{\varphi})]\rho+\gamma_{L}^{+}(\omega_{\varphi})\mathcal{D}[\mathbb{L}_{1}^{\dagger}(\omega_{\varphi})]\rho\\
    &\,\,\,+{\sum}_\varphi\gamma_{R}^{-}(\omega_{\varphi})\mathcal{D}[\mathbb{L}_{2}^{}(\omega_{\varphi})]\rho+\gamma_{R}^{+}(\omega_{\varphi})\mathcal{D}[\mathbb{L}_{2}^{\dagger}(\omega_{\varphi})]\rho,
\end{split}
\end{equation}
where $\varphi\in\{+,-\}$ denotes the corresponding eigenfrequency of the two coupled qubits $\omega_\pm^{}=\omega\pm g$, $\gamma_{\lambda}^{-}(\omega_{\varphi})=\gamma_{\lambda}^{}[1+n_{\lambda}^{}(\omega_{\varphi})]$, $\gamma_{\lambda}^{+}(\omega_{\varphi})=\gamma_{\lambda}^{}n_{\lambda}^{}(\omega_{\varphi})$, and $\lambda=L, R$ being the bath label. The system-bath interaction can only induce transitions between states with an energy difference of $\hbar(\omega \pm g)$, since the global jump operators~\cite{Global,Hofer}, known as eigenoperators, are given by $\mathbb{L}_{j}^{}(\omega_{\pm}^{})=|0\rangle \langle 0|\sigma_{j}^{-}|\omega_{\pm}^{}\rangle \langle \omega_{\pm}^{}|+|\omega_{\mp}^{}\rangle \langle\omega_{\mp}^{}|\sigma_{j}^{-}|2\rangle \langle 2|$.
These eigenoperators satisfy the commutation relation $[H_{\rm diag},\mathbb{L}_j^{}(\omega_\pm^{})]=-\omega_\pm^{} \mathbb{L}_j^{}(\omega_\pm^{})$, where $H_{\rm diag}$ is the total subsystem Hamiltonian in the diagonal basis.

To get the heat current in the global approach, we use the GME~(\ref{Lindblad_GME}) and calculate the explicit density matrix in the steady state as in~\cite{NJP_Geraldine_2020}, or the expectation values as in Appendix~\ref{appendix:a}, both paths yield to the following expression:
\begin{equation}\label{global_heat_current}
 J_{\rm global}^{(2)}=J_{\rm SS}^{}(\omega_{+}^{})+J_{\rm SS}^{}(\omega_{-}^{}),
\end{equation}
where
 \begin{equation}
     J_{\rm SS}^{}(\omega_{\pm})=\frac{ \gamma_{L}^{}\gamma_{R}^{}\,\omega_{\pm}^{}\big[n_{L}^{}(\omega_{\pm}^{})-n_{R}^{}(\omega_{\pm}^{})\big]}{\gamma_L^{}\coth\left(\frac{\hbar\omega_\pm^{}}{2k_B^{}T_L^{}}\right)+\gamma_R^{}\coth\left(\frac{\hbar\omega_\pm^{}}{2k_B^{}T_R^{}}\right)}.
 \end{equation} 
Contrary to Eq.~(\ref{general_heat_2}) obtained in the local approach, it is not possible to express $J_{\rm global}^{(2)}$ simply as a function of $J_{\rm SS}^{(1)}$. Instead, it splits into contributions associated with the eigenfrequencies $\omega_{\pm}$. By replacing $J^{(2)\rightarrow}_{\rm global}$ and $J^{(2)\leftarrow}_{\rm global}$ in the contrast definition, and by taking $T_{R} \to 0$ we get  
\begin{equation}\label{contr_strong}
    C_{\rm global}=\left|\frac{\gamma_{L}^{}-\gamma_{R}^{}}{\gamma_{L}^{}+\gamma_{R}^{}} \right|\left[1+\frac{A^{+}+A^{-}}{A^{+}n_{L}^{}(\omega_{+}^{})+A^{-}n_{L}^{}(\omega_{-}^{})}\right]^{-1},
\end{equation} 
where
\begin{align}
    A^{\pm}&=\!\frac{\omega_{\pm}^{}n_{L}^{}(\omega_{\pm}^{})}{\left[\frac{\gamma_{R}^{}}{\gamma_{L}^{}}\!+\!\coth{\left(\frac{\hbar\omega_{\pm}}{2k_B^{}T_{L}}\right)}\right]\left[\frac{\gamma_{L}^{}}{\gamma_{R}^{}}\!+\!\coth{\left(\frac{\hbar\omega_{\pm}}{2k_B^{}T_{R}}\right)}\right]}   .
\end{align}

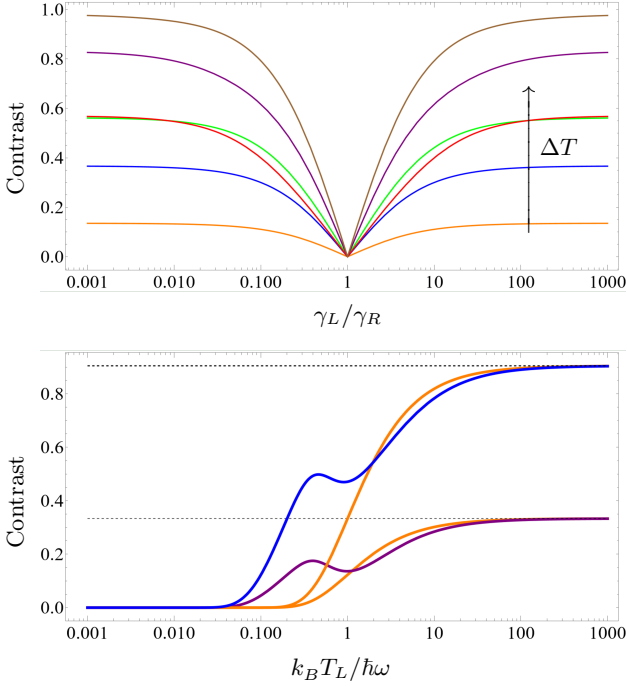
\begin{figure}[t]
\centering
\begin{tikzpicture} 
  \node (img1) {\includegraphics[width=.9\columnwidth]{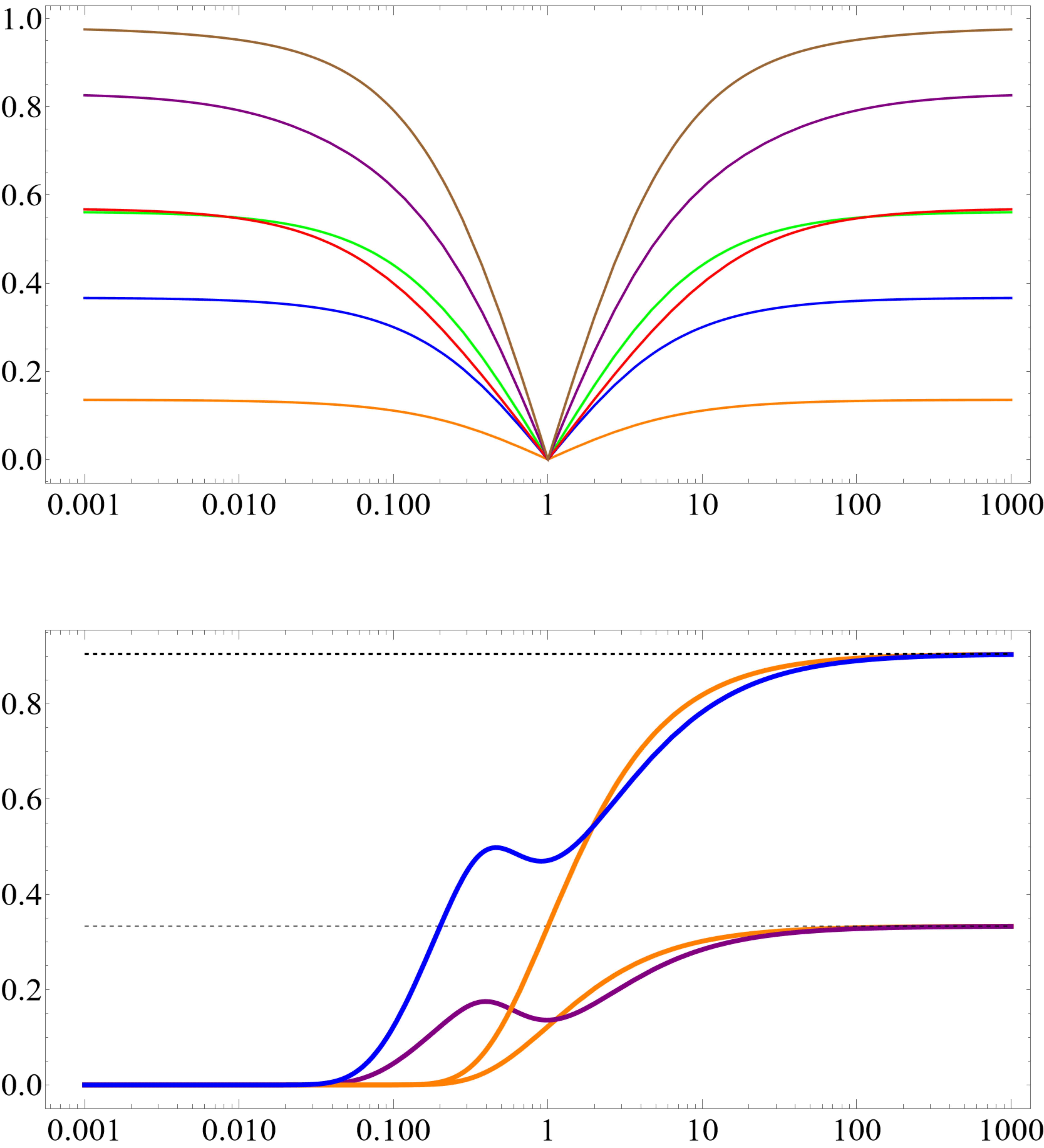}};
  \node[above=of img1, node distance=0cm, yshift=-10.3cm,xshift=0.1cm] {$k_{B}^{}T_L/\hbar\omega$};
  \node[above=of img1, node distance=0cm, yshift=-5.6cm,xshift=0.2cm] {$\gamma_{L}/\gamma_{R}$};
    \node[above=of img1, node distance=0cm,rotate=90, yshift=-2.8cm,xshift=-3.25cm] {$\overrightarrow{\hspace{2cm}}$};
    \node[above=of img1, node distance=0cm,yshift=-3.3cm,xshift=+3.0cm] {$\Delta T$};
  \node[above=of img1, node distance=0cm,rotate=90, yshift=4cm,xshift=-3.1cm] {{\color{black}Contrast}};
  \node[above=of img1, node distance=0cm,rotate=90, yshift=4cm,xshift=-7.8cm] {{\color{black}Contrast}};
\end{tikzpicture}
\caption{Thermal rectification. $C_{\rm global}$ as a function of $\gamma_L^{}/\gamma_R^{}$ (upper panel) for an increasing temperature bias $\Delta T$. Scaled bath temperatures are set as $k_{B}^{}T_{L}/\hbar \omega = 0.1, 0.2, 0.5, 1, 5, 50$ (from down to up). $C_{\rm global}$ as a function of the temperature (lower panel) for $\gamma_{L}^{}=0.5\gamma_{R}^{}$ (blue line) and $\gamma_{L}^{}=0.05\gamma_{R}^{}$ (purple line). The blue/purple line is the contrast $C_{\rm global}$ [Eq.~(\ref{contr_strong})] of two strongly ($g\geq \gamma_\lambda$) interacting  qubits. For comparison, the orange line is also the contrast, but for the case where the central system is one single qubit, $C_{\rm SS}^{(1)}$, see Eq.~(\ref{CBQB}). The parameters are $g=0.8\omega$,  $\omega=1$, $\gamma_{R}=0.02$ and $\gamma_{L}=0.01$ (purple line), and $\gamma_{L}=0.001$ (orange line).} 
\label{contr_strong_fig}
\end{figure}

\begin{figure}[t]
\centering
\begin{tikzpicture} 
  \node (img1) {\includegraphics[width=.9\columnwidth]{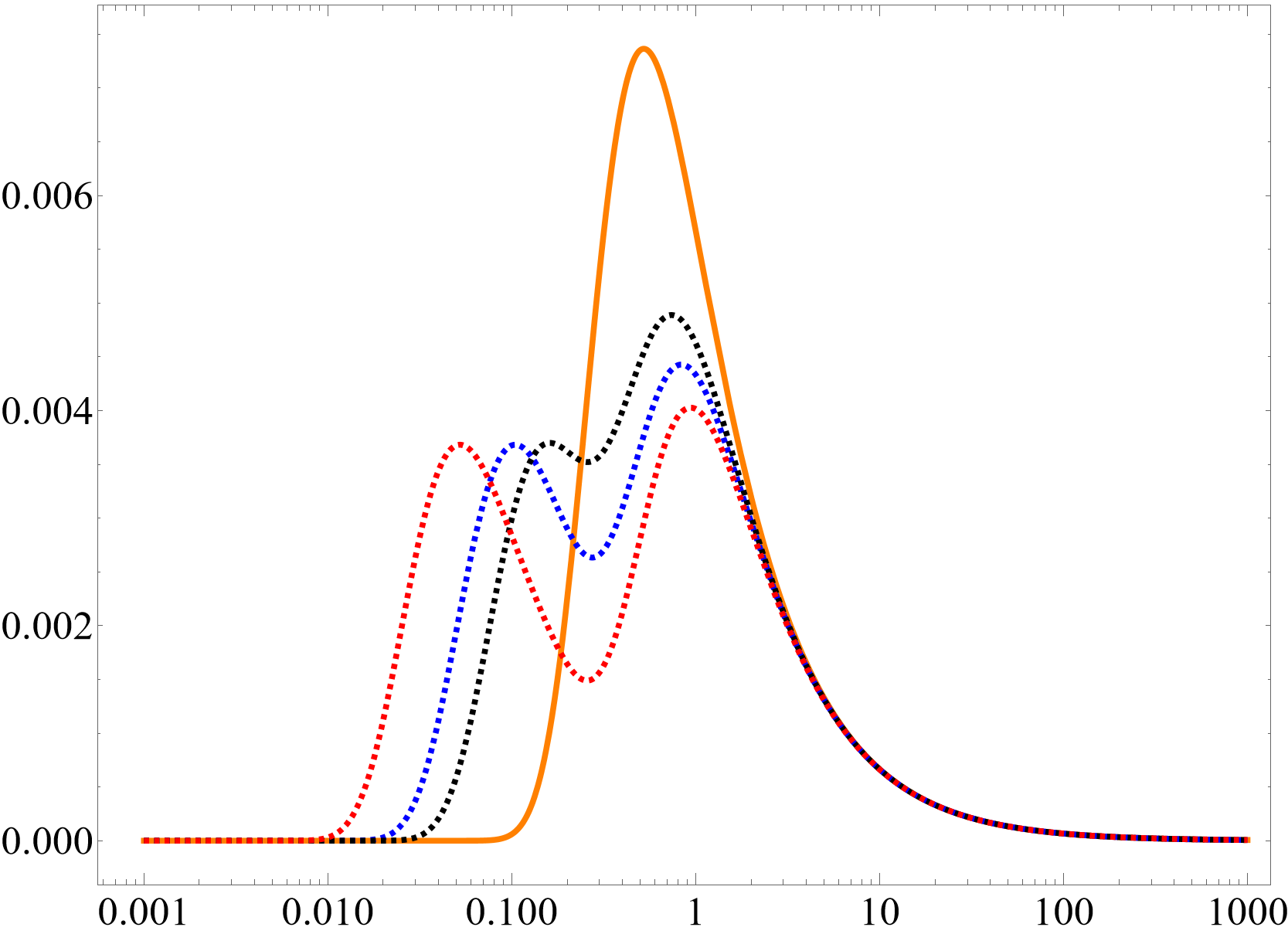}};
  \node[above=of img1, node distance=0cm, yshift=-7.5cm,xshift=0.1cm] {$k_{B}T/\hbar\omega$};
  \node[above=of img1, node distance=0cm,rotate=90, yshift=4cm,xshift=-3.8cm] {{\color{black}Thermal conductance}};
\end{tikzpicture}
\caption{Linear thermal conductance for one qubit (solid orange line) and two coupled qubits (dashed lines) in the strong coupling regime $\kappa_{\rm global}^{(2)}$. The inter-coupling constant, in decreasing order, is $g=0.9\omega$ (red dashed line), $g=0.8\omega$ (blue dashed line), and $g=0.7\omega$ (black dashed line). In all cases $\gamma_L^{}=0.01\omega$ and $\gamma_R^{}=0.02\omega$. }
\label{conduc_global_fig}
\end{figure}

In Fig.~\ref{contr_strong_fig}, we plot the contrast $C_{\rm global}$ as a function of $\gamma_L^{}/\gamma_R^{}$ (upper panel) for increasing temperature bias. Scaled bath temperatures are set as $k_{B}^{}T_{L}/\hbar \omega =0.1$ (orange line), $0.2$ (blue line), $0.5$ (green line), $1$ (red line), $5$ (purple line) and $50$ (brown line). The lower panel of Fig.~\ref{contr_strong_fig} shows $C_{\rm global}$ as a function of $T_L$ for two different ratios of the bath couplings, $\gamma_{L}^{}=0.5\gamma_R^{}$ (purple line) and $\gamma_{L}^{}=0.05\gamma_R^{}$ (blue line). We compare them with their corresponding $C_{\rm SS}^{(1)}$ of Eq.~(\ref{CBQB}) (orange lines). It is clear from Fig.~\ref{contr_strong_fig} that in the high-temperature limit $C_{\rm global}\rightarrow C_{\rm SS}^{(1)} $. Considering $k_B^{}T_{L}/\hbar\omega\gg1$, Eq.~(\ref{contr_strong}) approximates to $C_{\rm global} \approx|(\gamma_L^{}-\gamma_R^{})/(\gamma_L^{}+\gamma_R^{})|$, which coincides with the expression of $ C_{\rm SS}^{(1)} $ at high temperatures; see gray dotted lines. Thus, thermal rectification for $C_{\rm global}$ and $C_{\rm SS}^{(2)}$ coincide with $ C_{\rm SS}^{(1)} $ in the high-temperature regime. Notice that there are temperature ranges where $C_{\rm SS}^{(1)}\approx 0$, but the QTM under strong coupling still exhibits rectification. In contrast to the monotonic behavior of $C_{\rm SS}^{(1)}$, we observe a local maximum of $C_{\rm global}$ in a region where $C_{\rm SS}^{(1)}$ is small. There are also temperature values where $C_{\rm SS}^{(1)}$ slightly exceeds $C_{\rm global}$, indicating another difference from the local approach, where $C_{\rm SS}^{(1)}=C_{\rm SS}^{(2)}$ at any temperature. Note that while Eq.~(\ref{CBQB}) does not depend on $g$,  Eq.~(\ref{contr_strong}) does. 

The thermal conductance in the global approach, $\kappa_{\rm global}^{(2)}\equiv\lim_{\Delta T \to 0}^{}{J_{\rm global}^{(2)}}/\Delta T$, is 
\begin{align}
    \kappa_{\rm global}^{(2)}&=\frac{\gamma_{L}^{}\gamma_{R}^{}}{\gamma_{L}+\gamma_{R}}\left[\frac{\omega_{-}^{2}{\rm csch}\left(\beta\omega_{-}^{}\right)}{2k_B^{} T^{2}}+\frac{\omega_{+}^{2}{\rm csch}\left(\beta\omega_{+}^{}\right)}{2k_B^{} T^{2}}\right],
\end{align}
where $\beta=(k_B^{}T)^{-1}$. As with the heat current, $\kappa_{\rm global}^{(2)}$ cannot be written in terms of $\kappa_{\rm SS}^{(1)}$; see Eq.~(\ref{kappa_ss_2}). Instead, it can be viewed as the sum of two thermal conductances associated with two independent qubits of frequency $\omega_{\pm}$. Interestingly, we find that for large values of $g$, the typical single peak of thermal conductance splits into two asymmetric peaks, as shown in Fig.~\ref{conduc_global_fig}. In Appendix~\ref{appendix:f}, we show that the location of these peaks is close to $T\sim \omega_\pm$/2, and the separation between them increases with $g$. The full width at half maximum of each peak is approximately $3\omega_{\pm}/2$. This peculiar double-peak structure resembles the vacuum Rabi splitting in quantum optics, where the separation between the peaks is $2g$ \cite{Knight, Scully, Bishop}.

It is clear that with the coupling strength, $g$, we can conveniently tailor the quantum heat transport. Unlike the weak coupling regime, where $\kappa_{\rm SS}^{(2)}$ is bounded by the single peak structure of $\kappa_{\rm SS}^{(1)}$, there are temperature ranges in the strong coupling regime where $\kappa_{\rm SS}^{(1)}$ vanishes. At the same time, $\kappa_{\rm global}^{(2)}$ can be substantially different from zero; see the low-temperature regime of Fig.~\ref{conduc_global_fig}. 
The asymmetry of the peaks is because each eigenmode has a different energy gap, given by $\omega_\pm^{}=\omega\pm g$. Therefore, large (small) energy gaps allow for higher (lower) values of heat transport; this makes our QTM acts, in addition, as a heat valve~\cite{Ronzani2018}. These analytical results are consistent with recent numerical optimization findings, which highlight the superiority of strong coupling over weak coupling in enhancing heat rectification and conduction~\cite{PRR_Haack_2023}.


\section{Conclusions}\label{sec:conc}

We derived a general expression for the heat current [Eq.~(\ref{general_heat_2})] in the weak inter-subsystem coupling case, known as the local approach of open quantum systems. Our results allow us to analyze not only the specific system studied in this work but also other combinations of statistics in the baths and the subsystems. Specifically, for oscillator baths and qubits as subsystems, we have shown that the N-qubits chain, the two-qubit setup, and the one-qubit setup exhibit the same heat rectification behavior. This is because the current for an N-qubit chain is the same as for two qubits, and both are proportional to the current for one qubit. A similar relation of proportionality has been found for the thermal conductance. There is no distinction of which setup we use for heat rectification. The selection of the setup depends on other desired characteristics. For instance, the two-qubit system is preferable for generating entanglement and having a compact setup. However, the one-qubit system is more suitable if the goal is to maximize thermal conductance.

We found no fundamental relation between heat rectification and entanglement generation but a practical one. In contrast, rectification was tied to a change in the entropy production rate. These results contribute to the ongoing discussion on the role of quantum coherence and quantum correlations in nanoscale thermodynamics. Besides highlighting conditions for thermal rectification, we show the emergence of negative differential thermal conductance (resistance), a result not commonly observed in the weak coupling regime, but it is an essential requirement for the operation of a thermal transistor. We demonstrate how the strong coupling leads to an asymmetric Rabi-type splitting in the thermal conductance, enhancing heat transport and inaccessible rectification in the weak coupling; see Figs.~\ref{contr_strong_fig} and \ref{conduc_global_fig}. By conveniently tailoring the quantum heat transport across the bridge, our system represents the simplest QTM that consumes incoherent resources and delivers entanglement while acting as a rectifier and heat valve, offering insights that may inspire future research in the design and optimization of quantum devices.


\acknowledgments
R.R.-A. thanks DGAPA-UNAM, M\'exico for support under Project No. IA104624. M.S. would like to express his gratitude to CONAHCyT, Mexico for his Scholarship. B\c{C} is partially supported by The Scientific and Technological Research Council of Turkey (TUBITAK) under Grant No. 121F246.

\onecolumngrid
\appendix

\section{Equations of motion}\label{appendix:a}

Using the fact that in the steady state $Q_{L}^{\rm SS}=-Q_{R}^{\rm SS}$, we can write the heat current, $J_{\rm SS}^{(2)}= Q_{L}^{\rm SS}-Q_{R}^{\rm SS}$, in terms of the expectation value $\langle s_{j}^{\dagger}s_{j}^{}\rangle_{\rm SS}^{}$, such that
\begin{align}
    J_{\rm SS}^{(2)}
    =2\omega\gamma_{L}^{}\big( n_{L}^{}-[1+(\epsilon_{L}^{}-\epsilon_{a}^{})n_{L}^{}]\langle s_{1}^{\dagger}s_{1}^{}\rangle_{\rm SS}^{}\big).\label{Jss_s1s1}
\end{align}
To find $\langle s_{1}^{\dagger}s_{1}^{}\rangle_{\rm SS}^{} $ and $\langle s_{2}^{\dagger}s_{2}^{}\rangle_{\rm SS}^{}$, we use the master equation~(\ref{m_equation}), and we obtain the following coupled equations of motion
\begin{align}
    \frac{d}{dt}\langle s_{1}^{\dagger}s_{1}^{}\rangle&=(-\gamma_{L}^{-}+\epsilon_{a}^{}\gamma_{L}^{+})\langle s_{1}^{\dagger}s_{1}^{}\rangle+\gamma_{L}^{+}+ig(\langle s_{1}^{}s_{2}^{\dagger}\rangle-\langle s_{1}^{\dagger}s_{2}^{}\rangle),\label{S1S1EqMot}\\
    \frac{d}{dt}\langle s_{2}^{\dagger}s_{2}^{}\rangle&=(-\gamma_{R}^{-}+\epsilon_{a}^{}\gamma_{R}^{+})\langle s_{2}^{\dagger}s_{2}^{}\rangle + \gamma_{R}^{+}-ig(\langle s_{1}^{}s_{2}^{\dagger}\rangle-\langle s_{1}^{\dagger}s_{2}^{}\rangle),\\
    \frac{d}{dt}\langle s_{1}^{}s_{2}^{\dagger}\rangle&=+ig(\langle s_{1}^{\dagger}s_{1}^{}\rangle-\langle s_{2}^{\dagger}s_{2}^{}\rangle)+\frac{1}{2}[(\gamma_{L}^{+}+\gamma_{R}^{+})\epsilon_{a}^{}-(\gamma_{L}^{-}+\gamma_{R}^{-})]\langle s_{1}^{}s_{2}^{\dagger}\rangle,\\
    \frac{d}{dt}\langle s_{1}^{\dagger}s_{2}^{}\rangle&=-ig(\langle s_{1}^{\dagger}s_{1}^{}\rangle-\langle s_{2}^{\dagger}s_{2}^{}\rangle)+\frac{1}{2}[(\gamma_{L}^{+}+\gamma_{R}^{+})\epsilon_{a}^{}-(\gamma_{L}^{-}+\gamma_{R}^{-})]\langle s_{1}^{\dagger}s_{2}^{}\rangle.
\end{align}
At the steady state, each time derivative in the above equations vanishes, thus  we get
\begin{eqnarray}\label{S1S1Average}
    \langle s_{1}^{\dagger}s_{1}^{}\rangle_{\rm SS}^{}=\frac{4(\gamma_{L}^{+}+\gamma_{R}^{+})g^{2}+\gamma_{L}^{+}(\gamma_{R}^{-}-\epsilon_{a}^{}\gamma_{R}^{+})[\gamma_{L}^{-}+\gamma_{R}^{-}-\epsilon_{a}^{}(\gamma_{L}^{+}+\gamma_{R}^{2})]}{[4g^{2}+(\gamma_{L}^{-}-\epsilon_{a}^{}\gamma_{L}^{+})(\gamma_{R}^{-}-\epsilon_{a}^{} \gamma_{R}^{+})][\gamma_{L}^{-}+\gamma_{R}^{-}-\epsilon_{a}^{}(\gamma_{L}^{+}+\gamma_{R}^{+})]},\\
    \langle s_{2}^{\dagger}s_{2}^{}\rangle_{\rm SS}^{}=\frac{4(\gamma_{L}^{+}+\gamma_{R}^{+})g^{2}+\gamma_{R}^{+}(\gamma_{L}^{-}-\epsilon_{a}^{}\gamma_{L}^{+})[\gamma_{L}^{-}+\gamma_{R}^{-}-\epsilon_{a}^{}(\gamma_{L}^{+}+\gamma_{R}^{2})]}{[4g^{2}+(\gamma_{L}^{-}-\epsilon_{a^{}}\gamma_{L}^{+})(\gamma_{R}^{-}-\epsilon_{a}^{} \gamma_{R}^{+})][\gamma_{L}^{-}+\gamma_{R}^{-}-\epsilon_{a}^{}(\gamma_{L}^{+}+\gamma_{R}^{+})]},\\
    \langle s_{1}^{}s_{2}^{\dagger}\rangle_{\rm SS}=\frac{2ig(\gamma_{L}^{+}\gamma_{R}^{-}-\gamma_{L}^{-}\gamma_{R}^{+})}{[4g^{2}+(\gamma_{L}^{-}-\epsilon_{a}^{}\gamma_{L}^{+})(\gamma_{R}^{-}-\epsilon_{a}^{}\gamma_{R}^{+})][\gamma_{L}^{-}+\gamma_{R}^{-}-\epsilon_{a}^{}(\gamma_{L}^{+}+\gamma_{R}^{+})]},
\end{eqnarray}
and $\langle s_{1}^{\dagger}s_{2}\rangle_{\rm SS}^{}=(\langle s_{1}s_{2}^{\dagger}\rangle_{\rm SS}^{})^*$.
By substituting (\ref{S1S1Average}) into (\ref{Jss_s1s1}) and explicitly expressing the corresponding dissipation rates $\gamma_\lambda^\pm$ of the LME (\ref{m_equation}), the heat current between the left and right reservoirs, mediated by a bridge of two coupled quantum systems, is given by Eq.~(\ref{general_heat_2}) for any value of $\epsilon_\lambda^{}$ and $\epsilon_a^{}$.

\section{The scaling factor $\alpha$}\label{appendix:b}

\begin{figure}[ht]
\centering
\includegraphics[width=.4\textwidth,keepaspectratio=true]{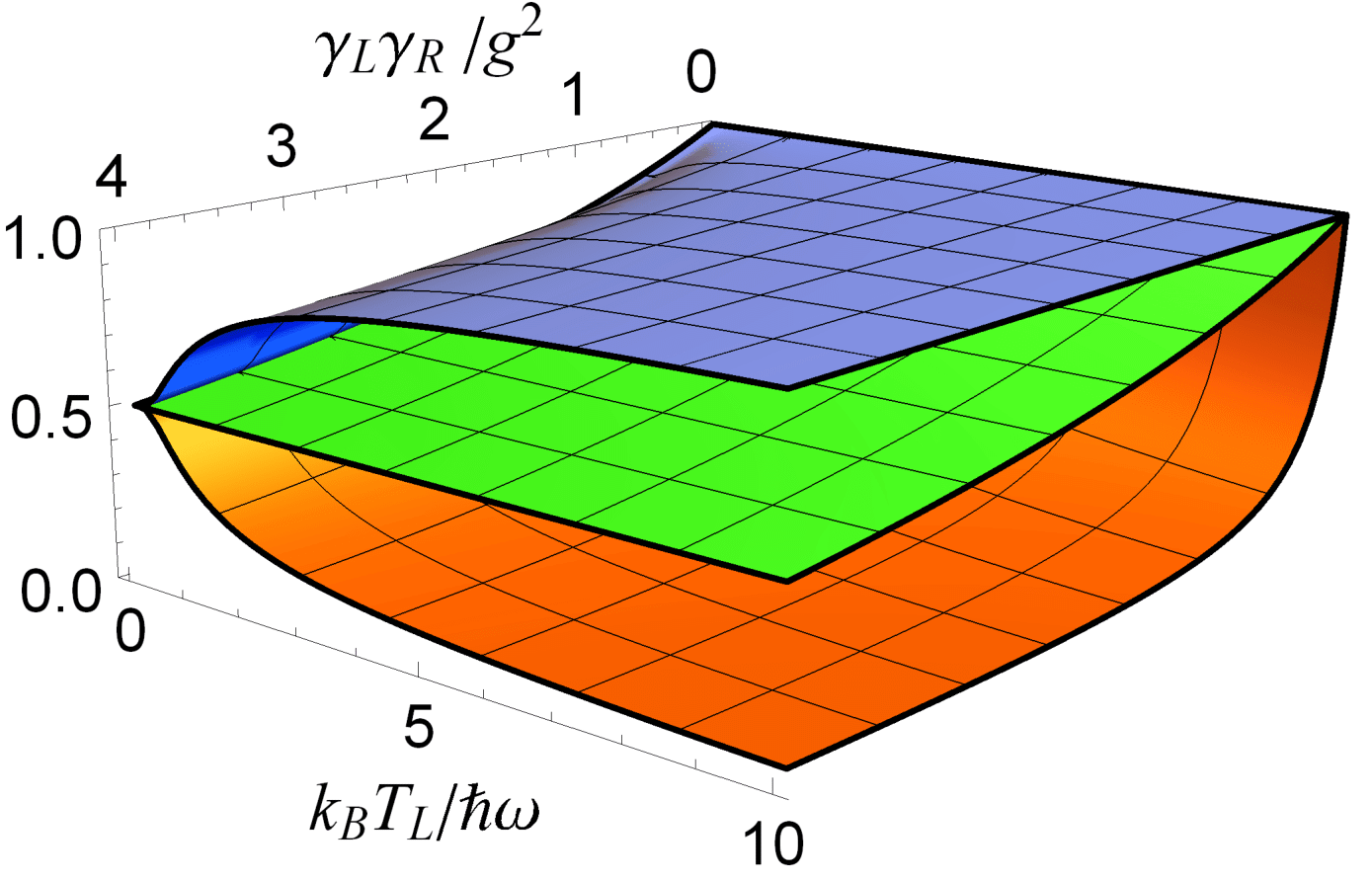}
\caption{Values of $\alpha_{0}^{}$ (green surface), $\alpha_{1}^{}$ (orange surface), and  $\alpha_{2}^{}$ (blue surface) as a function of the scaled temperature $k_B^{}T_L/\hbar\omega$ and $\gamma_L^{}\gamma_R^{}/g^{2}$.}
\label{alpha_figure}
\end{figure}

In the low-temperature limit of the right (cold) bath, where $k_B^{}T_R^{}/\hbar\omega\ll 1$ and $n_R^{}\approx 0$, the scaling factor $\alpha$ from Eq.~(\ref{alpha_factor}) becomes independent of $\epsilon_R^{}$. Thus, $\alpha$ is determined solely by the difference $\epsilon_L^{}-\epsilon_a^{}$, leading to three different expression for $\alpha$:
\begin{eqnarray}
\alpha_{0}^{}&\equiv&\Big[1+\frac{\gamma_L^{}\gamma_R^{}}{4g^2}\Big]^{-1}\hspace{1.8cm}\qquad\quad\,\,\, {\rm if}\,\, \epsilon_L^{}-\epsilon_a^{}=0,\label{appa}
\\
\alpha_{1}^{}&\equiv&\Big[1+\frac{\gamma_{L}^{}\gamma_{R}^{}}{4g^2}\coth{\Big(\frac{\hbar\omega}{2k_B^{}T_L}\Big)}\Big]^{-1}\qquad {\rm if}\,\, \epsilon_L^{}-\epsilon_a^{}=2,\quad \label{appb}
\\
    \alpha_{2}^{}&\equiv&\Big[1+\frac{\gamma_{L}^{}\gamma_{R}^{}}{4g^2}\tanh{\Big(\frac{\hbar\omega}{2k_B^{}T_L}\Big)}\Big]^{-1}\qquad {\rm if}\,\, \epsilon_L^{}-\epsilon_a^{}=-2.\quad \label{appc}
\end{eqnarray}
From Fig.~\ref{alpha_figure} we can see that $\alpha_{2}^{}\geq \alpha_{0}^{} \geq \alpha_{1}^{}$, where the values of $\alpha_i$ decrease whenever $x\equiv \gamma_L\gamma_R/4g^2$ increases. Therefore, according to the baths and subsystems statistics, $J_{\rm SS}^{(2)}$ is closer to $J_{\rm SS}^{(1)}$. 

\section{Critical heat current}\label{appendix:c}
Reference~\cite{NJP_Geraldine_2020} demonstrated that a minimum heat current is required to generate entanglement in a setup like the one we show in Fig.~\ref{fig_setup_two_qubit}. Here, we use the concurrence for an X-type density matrix \cite{EstadoX, EstadoX1}  to easily obtain the condition for such a minimum current. The density matrix of the two-coupled qubits in the steady state and in our notation is~\cite{NJP_Geraldine_2020}
\begin{equation}\label{Density_Matrix}
\rho_{\rm SS}^{}= \begin{pmatrix}
p_{1}^{} & 0 & 0 & 0 \\
0 & p_{2}^{} & \mathcal{C} & 0  \\
0 & \mathcal{C}^{*} & p_{3}^{} & 0 \\
0 & 0 & 0 & p_{4}^{}
\end{pmatrix},  
\end{equation}
where $p_{i}^{}$ with $i=1,2,3,4$ are the populations and $\mathcal{C}$ denotes the off-diagonal elements, namely coherence given by  $\mathcal{C} {\rm}=\langle s_{1}^{}s_{2}^{\dagger}-s_{1}^{\dagger}s_{2}^{}\rangle/2$. 

To quantify the entanglement content, we choose the well-known measure of concurrence defined by $C_{\rm conc}(\rho)= \rm max\{0,\Lambda_{1}-\Lambda_{2}-\Lambda_{3}-\Lambda_{4}\}$, where $0 \leq C_{\rm conc}(\rho) \leq 1$~\cite{Concurrence1,Concurrence2,Concurrence3}. $C_{\rm conc}(\rho)=0$ indicates zero entanglement and $C_{\rm conc}(\rho)=1$ the maximal pure state entanglement. $\Lambda_{i}$ are square roots of eigenvalues (in decreasing order) of the matrix $ R\equiv\rho (\sigma_{y} \otimes \sigma_{y}) \rho^{*} (\sigma_{y} \otimes \sigma_{y})$, where $\sigma_{y}$ is the Pauli matrix, and $\rho^{*}$ denotes complex conjugation of the original two-qubit density operator, i.e., $\rho_{\rm SS}^{}$. We note that~(\ref{Density_Matrix}) has the X-type form, which concurrence has been studied before~\cite{EstadoX,EstadoX1}. The eigenvalues of $R$ are $\Lambda_{1,4}=\big|\sqrt{p_{2}^{}p_{3}^{}}\pm|\mathcal{C}|\big|$ and $\Lambda_{2,3}=\sqrt{p_{1}^{}p_{4}^{}}$. The maximum eigenvalue is $\Lambda _{1} =\sqrt{p_{2}^{}p_{3}^{}} + |\mathcal{C}|$ and the concurrence is $ C_{\rm conc}(\rho)=2\,{\rm max}\{0,|\mathcal{C}|-\sqrt{p_{1}^{}p_{4}^{}}\}$, which leads to the following condition for the presence of entanglement between the two subsystems connecting the hot and cold baths $|\mathcal{C}|^{2}>p_{1}p_{4}$. In Sec.~\ref{heat_subsection} we find that $ \mathcal{C}_{\rm SS}^{}=\frac{i}{4g\omega}J_{\rm SS}^{(2)}$, therefore $J_{\rm SS}^{(2)}>4g\omega \sqrt{p_{1}^{}p_{4}^{}}\equiv J_{c}$, where $J_{c}$ denotes the critical heat current to generate entanglement in the steady state.

Solving the master equation~(\ref{m_equation}) to get the matrix elements (\ref{Density_Matrix}), one finds that~\cite{NJP_Geraldine_2020}
\begin{align}
    p_{1}^{}=\frac{4g^{2}(\gamma_{L}^{+}+\gamma_{R}^{+})^{2}+\gamma_{L}^{+}\gamma_{R}^{+}\Gamma^{2}}{(4g^{2}+\Gamma_{L}\Gamma_{R})\Gamma^{2}},
    \qquad
    p_{4}^{}=\frac{4g^{2}(\gamma_{L}^{-}+\gamma_{R}^{-})^{2}+\gamma_{L}^{-}\gamma_{R}^{-}\Gamma^{2}}{(4g^{2}+\Gamma_{L}\Gamma_{R})\Gamma^{2}},
\end{align}
where $\Gamma=\gamma_{L}^{+}+\gamma_{R}^{+}+\gamma_{L}^{-}+\gamma_{R}^{-}$, $\Gamma_{\lambda}=\gamma_{\lambda}^{-}+\gamma_{\lambda}^{+}$, and $\gamma_{\lambda}^{\pm}$ defined as in Sec.~\ref{sec:model} with $\lambda=L,R$. Finally, one obtains~\cite{NJP_Geraldine_2020}
\begin{align}\label{Critical_Heat_Current}
    J_{c}=\frac{8g^{2}\omega}{\chi}\bigg[4g^{2}(\gamma_{L}^{+}+\gamma_{R}^{+})^{2}(\gamma_{L}^{-}+\gamma_{R}^{-})^{2}+\Gamma^{2}(\gamma_{L}^{-}\gamma_{R}^{-}(\gamma_{L}^{+}+\gamma_{R}^{+})^{2}
    +\gamma_{L}^{+}\gamma_{R}^{+}(\gamma_{L}^{-}+\gamma_{R}^{-})^{2})+\gamma_{L}^{+}\gamma_{L}^{-}\gamma_{R}^{+}\gamma_{R}^{-}\frac{\Gamma^{4}}{4g}\bigg]^{1/2},
\end{align}
where $\chi=(4g^{2}+\Gamma_{L}\Gamma_{R})\Gamma^{2}$. Notice that $J_c$ is a general expression because $\gamma_{\lambda}^{\pm}$ lets us choose the statistic of the baths.

In the absence of heat rectification due to symmetric bath coupling, i.e., $\gamma_{L}^{}=\gamma_{R}^{}\equiv\gamma$, it is still possible to calculate the minimum heat current required to generate entanglement. Specifically, by applying (\ref{Critical_Heat_Current}) and  taking the limit $T_{R} \to 0$, we obtain:
\begin{equation}
    J_{c}=2n_{L}^{}\omega\alpha_{1}\frac{[4g^{2}(2+n_{L}^{})^{2}+\gamma^{2}[\coth{(\hbar\omega/2k_B^{}T_L)}+1]^{2}(1+n_{L}^{})]^{1/2}}{[\coth{(\hbar\omega/2k_B^{}T_L)}+1]^{2}}.
\end{equation} 

\section{Entropy production rate}\label{appendix:d}

Here, we compute the entropy production rate, which is defined as \cite{NJP_Chiara_2018,Kosloff,Landi_RMP_2021}
\begin{equation}\label{entropy_pro_def}
      \Pi=\frac{d S}{d t}-\sum_{\lambda} \beta_{\lambda}^{}J_{\lambda}^{},
\end{equation}
where $S$ is the von Neumann entropy of the two coupled subsystems and $J_{\lambda}$ is the heat current coming from the bath $\lambda$. At the steady state, $dS_{\rm SS}^{}/dt=0$, and (\ref{entropy_pro_def}) simplifies to $\Pi_{\rm SS}^{}=-\big(\beta_L^{} J_L^{}+\beta_R^{} J_R^{}\big)=(\beta_R^{}-\beta_L^{})J_{\rm SS}^{(2)}$. Since $T_L^{}>T_R^{}$, $\Pi_{\rm SS}^{}\ge 0$, which is the mathematical statement of the second law of thermodynamics. In our case, $\Pi_{\rm SS}^{}$ results in a finite and constant vale, the signature of a nonequilibrium steady state (NESS)~\cite{Landi_RMP_2021}. Comparing the entropy production rate for the case $J_{\rm SS}^{(2)\rightarrow}$, denoted by $\Pi_{\rm SS}^{\rightarrow}$, with the entropy production of the process $J_{\rm SS}^{(2)\leftarrow}$, denoted by $\Pi_{\rm SS}^{\leftarrow}$, yields
\begin{equation}\label{contrast_entropyprod}
    \frac{|\Pi_{\rm SS}^{\rightarrow}|}{|\Pi_{\rm SS}^{\leftarrow}|}=\frac{|J_{\rm SS}^{(2)\rightarrow}|}{|J_{\rm SS}^{(2)\leftarrow}|}=\mathcal{R},
\end{equation}
which is precisely the rectification coefficient~\cite{Exp_Rect5}. The above equation implies that the entropy production rate in both processes, $\Pi_{\rm SS}^{\rightarrow}$ and $\Pi_{\rm SS}^{\leftarrow}$, will be the same as long as there is no thermal rectification, i.e., $\mathcal{R}=1$. Due to the generality of (\ref{contrast_entropyprod}), an asymmetric entropy production rate implies the existence of thermal rectification in the QTM and vice versa. A similar relationship can be found for the entropy production rate as (\ref{contrast_entropyprod}) for $J_{\rm SS}^{(N)}$. Therefore, the entropy production rate for two qubits (harmonic oscillators) is the same as for an N-qubit (harmonic) chain.

\section{Peaks location of the thermal conductance in the global approach}
\label{appendix:f}
\begin{figure}[ht]
\centering
\includegraphics[width=.45\textwidth,keepaspectratio=true]{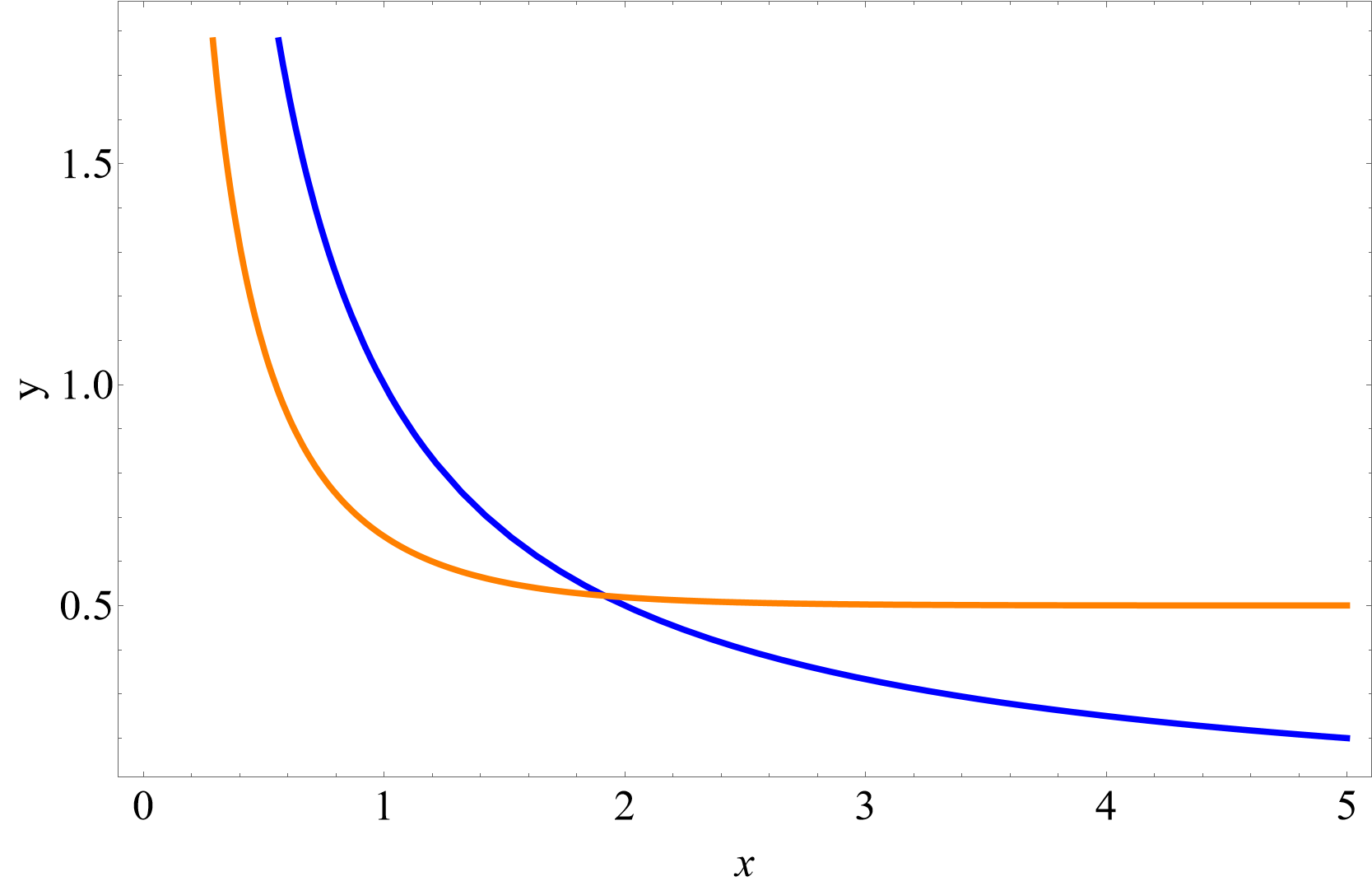}
\caption{Graph of $1/x$ (blue line) and $(\coth{x})/2$ (orange line).}
\label{transcendental}
\end{figure}

To determine the location of the peaks in $ \kappa_{\rm global}^{(2)}$ shown in Fig.~\ref{conduc_global_fig}, we solve the equation $d\kappa_{\rm global}^{(2)}/dT=0 $ for $T$:
\begin{align}
    \frac{d\kappa_{\rm global}^{(2)}}{d T}=-\omega_{-}^{2}{\rm csch}\left(\frac{\omega_{-}}{T}\right)-\omega_{+}^{2}{\rm csch}\left(\frac{\omega_{+}}{T}\right)+\frac{\omega_{-}^{3}\coth{\left(\frac{\omega_{-}}{T}\right)}{\rm csch}\left(\frac{\omega_{-}}{T}\right)+\omega_{+}^{3}\coth{\left(\frac{\omega_{+}}{T}\right)}{\rm csch}\left(\frac{\omega_{+}}{T}\right)}{2T}=0.
\end{align}
This is a transcendental equation where we make $k_B^{}=1$ for simplicity. To get an approximate solution, we notice that there are terms with $\omega_{-}$ and terms with $\omega_{+}$, so the peaks are near the local maxima of the functions related to these terms. We split the equation into two equations according to this 
\begin{align}
    -\omega_{-}^{2}{\rm csch}\left(\frac{\omega_{-}}{T}\right)+\frac{\omega_{-}^{3}}{2T}\coth{\left(\frac{\omega_{-}}{T}\right)}{\rm csch}\left(\frac{\omega_{-}}{T}\right)&=0,\\
    -\omega_{+}^{2}{\rm csch}\left(\frac{\omega_{+}}{T}\right)+\frac{\omega_{+}^{3}}{2T}\coth{\left(\frac{\omega_{+}}{T}\right)}{\rm csch}\left(\frac{\omega_{+}}{T}\right)&=0.
\end{align}
After some algebra, the above equations reduce to
\begin{align}
    \frac{T}{\omega_{-}}&=\frac{1}{2}\coth{\left(\frac{\omega_{-}}{T}\right)},\\
    \frac{T}{\omega_{+}}&=\frac{1}{2}\coth{\left(\frac{\omega_{+}}{T}\right)}.
\end{align}
Defining the variable $x\equiv\omega_\pm/T$, we plot the functions $y=x^{-1}$ and $y=\frac{1}{2}\coth{x}$ in Fig.~\ref{transcendental}. The intersection of these functions determines the peak location and can be approximated as $x\approx 2$, yielding $T_{\pm}\simeq \omega_{\pm}/2$.
\twocolumngrid

\end{document}